\documentclass{tcibook}
\usepackage{fancyhea}
\usepackage{work}
\usepackage{bm}       
\usepackage{graphicx}
\usepackage{multirow}
\usepackage{lineno}
\usepackage{threeparttable}
\usepackage[normalem]{ulem}
\usepackage{color}
\usepackage[colorlinks = true,
            linkcolor = blue,
            urlcolor  = blue,
            citecolor = blue,
            anchorcolor = blue]{hyperref}

\def\beq{\begin{equation}}
\def\eeq#1{\label{#1}\end{equation}}
\def\eeqn{\end{equation}}

\newenvironment{Eqnarray}%
   {\arraycolsep 0.14em\begin{eqnarray}}{\end{eqnarray}}
\def\beqa{\begin{Eqnarray}}
\def\eeqa#1{\label{#1}\end{Eqnarray}}
\def\eeqan{\end{Eqnarray}}

\let\bar=\overbar

\def\lsim{\mathrel{\raise.3ex\hbox{$<$\kern-.75em\lower1ex\hbox{$\sim$}}}}
\def\gsim{\mathrel{\raise.3ex\hbox{$>$\kern-.75em\lower1ex\hbox{$\sim$}}}}

\def\del{\partial}
\def\Dslash{\not{\hbox{\kern-4pt $D$}}}
\def\dslash{\not{\hbox{\kern-2pt $\del$}}}
\def\pslash{\not{\hbox{\kern-2pt $p$}}}
\def\ETmiss{\not{\hbox{\kern-4pt $E$}}_T}

\def\Dlr{\mathrel{\raise1.5ex\hbox{$\leftrightarrow$\kern-1em\lower1.5ex\hbox{$D$}}}}

\def\MSB{{\bar{M \kern -2pt S}}}
\def\msb{{\bar{\scriptsize M \kern -1pt S}}}

\def\drb{{\bar{\scriptsize D \kern -1pt R}}}

\begin{document}


\pagenumbering{roman}

\parindent=0pt
\parskip=8pt
\setlength{\evensidemargin}{0pt}
\setlength{\oddsidemargin}{0pt}
\setlength{\marginparsep}{0.0in}
\setlength{\marginparwidth}{0.0in}
\marginparpush=0pt


\pagenumbering{arabic}

\renewcommand{\chapname}{chap:intro_}
\renewcommand{\chapterdir}{.}
\renewcommand{\arraystretch}{1.25}
\addtolength{\arraycolsep}{-3pt}































\setcounter{chapter}{6} 

\chapter{Accelerator Frontier}

\vspace{1cm}
 
{\textbf{Frontier Conveners:} S.~Gourlay$^{1}$, T.~Raubenheimer$^{2}$, V.~Shiltsev$^{3}$}

{\textbf{Topical Group Conveners:} G. Arduini$^{4}$, R.~Assmann$^{5}$, C.~Barbier$^{6}$, M.~Bai$^{2}$, S.~Belomestnykh$^{3}$, S.~Bermudez$^{4}$,  A.~Faus-Golfe$^{7}$, J.~Galambos$^{6}$, C.~Geddes$^{1}$,  G.~Hoffstaetter$^{8}$, M.~Hogan$^{2}$,  Z.~Huang$^{2}$, M.~Lamont$^{4}$, D.~Li$^{1}$, S.~Lund$^{9}$, R.~Milner$^{10}$, P.~Musumeci$^{11}$, E.~Nanni$^{2}$,  M.~Palmer$^{12}$, N.~Pastrone$^{13}$,  F.~Pellemoine$^{3}$, E.~Prebys$^{14}$, Q.~Qin$^{15}$, G.~Sabbi$^{1}$, Y.-E.~Sun$^{16}$,  J.~Tang$^{17}$, A.~Valishev$^{3}$, H.~Weise$^{5}$, F.~Zimmermann$^{4}$, A.V.~Zlobin$^{3}$,  R.~Zwaska$^{3}$}

{\textbf{Contributors:} P.~Bhat$^{3}$, J.~Power$^{16}$, T.~Roser$^{12}$, D.~Stratakis$^{3}$}

\vspace{0.5cm}
{
$^{1}${\it Lawrence Berkeley National Laboratory, Berkeley, CA 94720, USA}\\
$^{2}${\it SLAC National Accelerator Laboratory, Menlo Park, CA 94025, USA}\\
$^{3}${\it Fermi National Accelerator Laboratory, Batavia, IL 60510, USA}\\
$^{4}${\it CERN, 1211 Meyrin, Switzerland}\\
$^{5}${\it Deutsches Elektronen-Synchrotron, 22607 Hamburg, Germany}\\
$^{6}${\it Oak Ridge National Laboratory, Oak Ridge, TN 37830, USA}\\
$^{7}${\it Université Paris-Saclay, CNRS/IN2P3, 91406 Orsay, France}\\
$^{8}${\it Cornell University, Ithaca, NY 14850, USA}\\
$^{9}${\it Michigan State University, East Lansing, MI 48824, USA}\\
$^{10}${\it MIT, Cambridge, MA 02139, USA}\\
$^{11}${\it UCLA, Los Angeles, CA 90095, USA}\\
$^{12}${\it Brookhaven National Laboratory, Upton, NY 11973, USA}\\
$^{13}${\it INFN Torino, 10127 Torino TO, Italy}\\
$^{14}${\it University of California, Davis, Davis, CA, 90095, USA}\\
$^{15}${\it ESRF, 38000 Grenoble, France}\\
$^{16}${\it Argonne National Laboratory, Lemont, IL 60439, USA}\\
$^{17}${\it IHEP, Bejing, 100039 China}\\
}

\newpage
  

For over half a century, high-energy accelerators have been a major enabling technology for particle and nuclear physics research as well as sources of X-rays for photon science research in material science, chemistry and biology. Particle accelerators for energy and intensity frontier research in high energy physics (HEP) continuously drive the accelerator community to invent ways to increase the energy and improve the performance of accelerators, reduce their cost, and make them more power efficient. The increasing size, cost and timescale required for modern and future accelerator-based HEP
projects arguably distinguish them as some of the most challenging scientific research endeavors and demand continuation of these efforts.
 In the meantime, the international accelerator community has demonstrated imagination and creativity in developing a plethora of future accelerator ideas and proposals. 

Major developments since the last  Snowmass/HEPAP P5 strategic planning exercise in 2013-2014 include start of the PIP-II proton linac; construction for the LBNF/DUNE neutrino program in the US; emergence of a number of projects for Higgs/EW physics such as FCC-ee at CERN, CEPC in China and C$^3$ and HELEN in the US; a significant reduction of activity related to linear collider projects (ILC in Japan and CLIC at CERN); and paradoxically, the end of the Muon Accelerator Program in the US and creation of the International Muon Collider Collaboration (IMCC) in Europe. The last decade saw several notable planning advancements, including the US DOE GARD Roadmaps, European Strategy for Particle Physics and the Accelerator R\&D Roadmap, EuPRAXIA, etc. 

In addition, since the last Snowmass meeting that took place in 2013 was shortly after the confirmation of the Higgs, the goals for the Energy Frontier have changed as a result of the LHC measurements.  While a Higgs/EW factory at 250 to 360 GeV is still the highest priority for the next large accelerator project, the motivation for a TeV or few TeV $e^+e^-$ collider has diminished.  Instead, the community is focused on a 10+ TeV (parton c.m.e) discovery collider that would follow the Higgs/EW Factory.  This is an important change that will refocus some of the accelerator R\&D programs. 

The technical maturity of proposed facilities ranges from shovel-ready to those that are still largely conceptual. Over 100 contributed papers have been submitted to the {\it Accelerator Frontier} of the US particle physics decadal community planning exercise, {\it Snowmass’2021}. These papers cover a broad spectrum of topics: beam physics and accelerator education, accelerators for neutrinos, colliders for Electroweak/Higgs studies and multi-TeV energies, accelerators for {\it Physics Beyond Colliders} and rare processes, advanced accelerator concepts, and accelerator technology for Radio Frequency cavities (RF), magnets, targets, and sources. 

In 2020-2022, extensive discussions and deliberations have taken place in corresponding topical working groups of the Snowmass Accelerator Frontier (AF) and in numerous joint meetings with other Frontiers, Snowmass-wide meetings, a series of Colloquium-style {\it Agoras}, cross-Frontier {\it Forums on muon and $e^+e^-$ colliders} and the collider {\it Implementation Task Force 
(ITF)}. The outcomes of these activities are summarized below.\\

{\bf Future facilities:} The accelerator community in the US and globally has a broad array of accelerator technologies and expertise that will be needed to design and construct any of the near-term HEP accelerator projects. P5 will need to prioritize what option(s) should be developed. \textbf{Planning of accelerator development and research should be aligned with the strategic planning for particle physics and should be part of the P5 prioritization process}. Accelerator experts can contribute to the US and international projects under consideration by providing top-down metrics for expected cost-scales and technology/timeline evaluations, following the ITF findings. 

Among possible actively discussed future facilities options are: 
\begin{itemize}
   \item A multi-MW beam power upgrade of the Fermilab proton accelerator complex that seems to be the highest priority for the neutrino program in the 2030s; corresponding accelerator technology and beam physics studies are needed to identify the most cost- and power-efficient solutions that  could be implemented in a timely fashion leading to breakthrough results of the DUNE neutrino program;
\item Several beam facilities for axion and Dark Matter (DM) searches are shown to have great potential for construction in the 2030s in terms of scientific output, cost and timeline, including PAR (a 1 GeV, 100 kW PIP-II Accumulator Ring); in general, we should focus on more efficient utilization of existing and upcoming facilities to explore dedicated or parasitic opportunities for rare process measurements - examples are the SLAC SRF electron linac, MWs of proton beam power potentially available after construction of the PIP-II SRF linac, spigots of the future multi-MW FNAL complex upgrade, and at CERN, a Forward Physics Facility at the LHC, etc; 
\item Possible future colliders - In addition to continuing support for the ILC that figured prominently in the last P5, there are several other approaches identified as both promising and potentially feasible, and call for further exploration and support: in the Higgs/EW sector - there is growing support for the FCC-ee at CERN and proposals of somewhat more advanced linear colliders in the US or elsewhere, such as $C^3$ and HELEN; 
\item At the energy frontier, discovery machines such as $O$(10 TeV c.m.e.) muon colliders have rapidly gained significant momentum. To be in a position for making decisions on collider projects viable for construction in the 2040s and beyond at the time of the next  Snowmass/P5, these concepts could be explored technically and documented in pre-CDR level reports by the end of this decade. 
\end{itemize}

The U.S. HEP accelerator R\&D portfolio presently contains no collider-specific scope.  This creates a gap in our knowledge-base and accelerator/technology capabilities.  It also limits our national aspiration for a leadership role in particle physics in that the US cannot lead or even contribute to proposals for accelerator-based HEP facilities.  To address the gap, the community has proposed that the U.S. establish a \textbf {national integrated R\&D program on future colliders} in the DOE Office of High Energy Physics (OHEP) to carry-out technology R\&D
and accelerator design for future collider concepts. This program would aim to enable synergistic engagement in projects proposed abroad (e.g. FCC, ILC, IMCC). It would support the development of design reports on collider options by the time of the next Snowmass and P5 (2029–2030), particularly for options that can be hosted in the US, and the creation of R\&D plans for the decades past 2030.  \textbf {Without such a program there may be few accelerator-based proposals for a future P5 to evaluate.}

\textbf{General Accelerator R\&D:}
In addition to the above focused proposed activity, general accelerator technology development is critical and needs additional attention and support to achieve the community aspirations by the end of this decade:
\begin{itemize}
    \item Novel high power targets should be developed to be able to accept multi-MW beams (up to 2.4 MW for the LBNF/DUNE Phase II, 4-8 MW for a future muon collider) - the task that requires dedicated high-power targetry R\&D facility(ies); development of efficient high intensity, high brightness $e^+$ sources is critical for most Higgs/EW factory colliders;
\item Energy-efficient SC and cold-NC RF cavities and structures need to demonstrate 70 MV/m and 70 to 150 MV/m gradients, respectively, which are needed for cost efficient compact Higgs factory linear colliders; also critical are exploration and testing of new materials with the potential of sustaining higher gradients and high $Q_0$ as well as development of efficient RF sources that could lead to significant cost reduction;
\item Conceptual breakthroughs in 12-20 T high-field dipole magnets, $O$(30 T) solenoids and $O$(1000 T/s) fast ramping magnets should demonstrate convincing proof of feasibility for high energy proton-proton and muon colliders as well as other energy frontier collider schemes;
\item Advanced wakefield accelerator concepts should strive toward demonstration of collider quality beams, efficient drivers and staging, and development of self-consistent parameter sets for potential  colliders based on wakefield acceleration in plasma and structures (in close coordination with international programs such as the European Roadmap, EuPRAXIA, etc.); 
\item Finally, in accelerator and beam physics - the focus should be on experimental, computational and theoretical studies on acceleration and control of high intensity/high brightness beams,  high performance computer modeling and AI/ML approaches, and design integration and optimization. The program should also include the overall energy efficiency of future facilities and re-establish a program of beam physics research on general collider-related topics towards future $e^+e^-$ colliders and muon colliders. 
\end{itemize}

The above items call for a substantial increase in accelerator R\&D support. We suggest that corresponding discussions take place as part of the P5 deliberations and clearly formulate priorities that should be communicated to the OHEP and accelerator community. A series of follow-up  workshops should update and formalize the U.S. strategic goals and roadmaps for each of the GARD thrusts. Education and workforce training for particle accelerators is of critical importance for several DOE SC offices (HEP, NP, BES, FES,...) and OHEP should take a lead in organizing discussions and formulation of a corresponding vision across the Office of Science and NSF. Additional measures are needed to make the US competitive in education and attraction of accelerator talent, such as:
%
\begin{itemize}
\item Strengthening and expansion of education and training programs, enhancement of recruiting, and promotion of the field (e.g., via colloquia at universities); 
\item Creating an undergraduate level recruiting program structured to draw in women and underrepresented minorities (URM) that could be coordinated with the USPAS, and corresponding efforts at all career stages to support, include and retain them in the field; 
\item Strengthen and expand capabilities of the US accelerator beam test facilities to maintain their competitiveness with respect to worldwide capabilities.
\end{itemize}

\tableofcontents

\newpage

\section{Introduction: progress since the last P5 (2014), Snowmass'21 Accelerator Frontier activities} 

For almost a century, high-energy particle accelerators have played a key role in shaping modern nuclear and particle physics. They also support forefront material science and biology research \cite{shiltsev2020particle}. 
Some 140 large accelerators are in operation that support research and many are currently under construction -- the High Luminosity LHC upgrade, PIP-II, NICA, XFELs, EIC, ESS, FAIR, etc -- and will become operational in due time. We also see great progress toward future frontier facilities for neutrino and rare processes physics research,  Higgs factories (linear or circular), and multi-TeV  $pp, \mu\mu$ and $e^+e^-$ colliders -- all of which are in the focus of the Snowmass'21 discussions. 

\subsection{Snowmass process}

Snowmass is a particle physics community study that takes place in the US every 7-9 years (the last one  was in 2013).  Snowmass'21 strives to define the most important questions for the field and to identify promising opportunities to address them, see  https://snowmass21.org/. It provides an opportunity for the entire particle physics community to come together to identify and document a scientific vision for the future of particle physics in the U.S. and its international partners. Following in the footsteps of Snowmass, the HEPAP Particle Physics Project Prioritization Panel (P5) will take the scientific input from Snowmass'21 (final summaries to be available in September 2022)  and by the Spring of 2023 develop a strategic plan for U.S. particle physics that can be executed over a 10 year timescale, in the context of a 20-year global vision for the field. 

\subsection{2014 P5 recommendations}

Previous P5 plans were developed in 2014 \cite{2014P5} and contained several recommendations related to accelerators, all of which have generally been addressed and implemented. Those include: 
\begin{itemize}
\item {\it Recommendation 10: Complete the LHC phase-1 upgrades and
continue the strong collaboration in the LHC with the phase-2
(HL-LHC) upgrades of the accelerator and both general-purpose
experiments (ATLAS and CMS). The LHC upgrades constitute
our highest-priority near-term large project} -- the AUP-LHC project has started delivery of a few dozen large aperture high field Nb$_3$Sn IR quads for the HL-LHC; led by FNAL, BNL, and LBNL it is currently at the level of CD-3 and is on track to be completed by the LHC Long Shutdown 3 (LS3). 
    \item {\it Recommendation 11: Motivated by the strong scientific importance of the ILC and the recent initiative in Japan to host it,
the U.S. should engage in modest and appropriate levels of ILC
accelerator and detector design in areas where the U.S. can
contribute critical expertise. Consider higher levels of collaboration
if ILC proceeds.} --  significant effort on the ILC design, prototyping and testing culminated in a beam acceleration demonstration of the full ILC cryomodule with an acceleration gradient of 31.5 MV/m at the FNAL FAST facility \cite{broemmelsiek2018record}.
\item {\it Recommendation 14: Upgrade the Fermilab proton accelerator
complex to produce higher intensity beams. R\&D for the Proton
Improvement Plan II (PIP-II) should proceed immediately, followed
by construction, to provide proton beams of $\ge$1 MW by
the time of first operation of the new long-baseline neutrino
facility} -- the 800 MeV H$^-$ PIP-II SRF linac construction has started, the project aims to support 1.2 MW of proton power on the neutrino target for LBNF/DUNE, with first beam injection to the FNAL Booster in 2029;  international contributions support some 30\% of the project. 
\item {\it Recommendation 22: Complete the Mu2e and muon g-2
projects} -- the FNAL accelerator complex has been modified to include Muon Campus machines and new beamlines that delivered 8 GeV protons on target to generate copious amounts of muons for the g-2 experiment, exceeding previous (BNL) statistics 12-fold; the proton beamline for the Mu2e experiment was commissioned in 2022. 
\item {\it Recommendation 25: Reassess the Muon Accelerator Program
(MAP). Incorporate into the GARD program the MAP activities
that are of general importance to accelerator R\&D, and consult
with international partners on the early termination of MICE.} -- the US played a major part in the MICE ionization cooling experiment at RAL (UK) that  successfully demonstrated $O$(10\%) transverse emittance reduction (cooling) of 140 MeV/c muons \cite{mice2020demonstration}; the US MAP program was closed in 2016 just in time to witness creation of the CERN-led International Muon Collider Collaboration (IMCC) in Europe a few years later.
\item {\it Recommendation 24: Participate in global conceptual design
studies and critical path R\&D for future very high-energy proton-
proton colliders. Continue to play a leadership role in
superconducting magnet technology focused on the dual goals
of increasing performance and decreasing costs.} -- many US accelerator experts have contributed to the design of the 100 TeV c.m.e FCChh collider \cite{fcchh}; a 14.5 T $Nb_3Sn$ dipole has been developed by the US Magnet Development Program (MDP) and successfully tested in 2021 \cite{zlobin2021ieee}. 
\end{itemize}

\subsection{2015 GARD subpanel recommendations}

Two remaining 2014 P5 recommendations -- {\it Recommendation 23: Support the discipline of accelerator science through advanced accelerator facilities and through funding for university programs. Strengthen national laboratory-university R\&D partnerships, leveraging their diverse expertise and facilities.} {\it Recommendation 26: Pursue accelerator R\&D with high priority at levels consistent with budget constraints. Align the present
R\&D program with the P5 priorities and long-term vision, with
an appropriate balance among general R\&D, directed R\&D, and
accelerator test facilities and among short-, medium-, and long-term
efforts. Focus on outcomes and capabilities that will dramatically
improve cost effectiveness for mid-term and far-term
accelerators.} -- this recommendation was addressed by a follow-up HEPAP  Accelerator R\&D Subpanel in 2015\cite{GARD2015subpanel}. It was charged to identify the most promising accelerator research areas to support the advancement of HEP and recommended vigorous beam physics and accelerator research along five thrusts: accelerator and beam physics (ABP), RF acceleration technology (NC and SC RF), magnets and materials, advanced acceleration concepts (AAC), particle sources and targets. 

The US OHEP supported these programs and corresponding facilities operations with a total of roughly 90M\$ a year, distributed as estimated from FY19 as 36\% for AAC, 22\% for RF, 19\% for materials and magnets, 15\% for ABP, and less than 2\% for targets and sources. 

Most of the GARD thrusts have demonstrated impressive accomplishments since 2015, the most notable include (see more details below) construction and operation of the FACET-II User facility at SLAC (with unique 1nC, 1$\times$1$\times$1$\mu$m 10 GeV electron bunches available to several hundred users), numerous breakthroughs at the BELLA laser-plasma Wake Field Accelerator (WFA) facility at LBNL (such as electron acceleration up to 8 GeV over a mere 0.2 m of plasma and a proof-of-principle 0.1 GeV + 0.1 GeV staging demonstration), construction of and start of beam physics research at the IOTA Ring at FNAL (with recent demonstration of Optical Stochastic Cooling  of 100 MeV electrons), a functional 14.5 T Nb$_3$Sn magnet and 300 T/s HTS magnet prototypes and new advanced high-current density conductors employing artificial pinning centers; new surface processing techniques resulted in 35-55 MV/m gradients in SRF cavities with unprecedented quality factors (which have been successfully applied to the LCLS-II/HE project). The high power targetry research thrust is the only one among the approved GARD thrust areas that does not have a (very much needed) test facility yet. 

\subsection{Snowmass'21 Accelerator Frontier}

The Snowmass'21 activities are managed along the lines of ten "Frontiers": 
Energy Frontier (EF), Neutrino Physics Frontier (NF), Rare Processes and Precision Frontier (RPF), Cosmic Frontier(CF), Theory Frontier (TF),  Accelerator Frontier (AF),  Instrumentation Frontier (IF), Computational Frontier (CompF),  Underground Facilities (UF), and Community Engagement Frontier (CEF). The Snowmass Community Summer Study workshop took place in Seattle at the University of Washington on July 17-26 and represented the culmination of the various workshops and Town Hall meetings that took place during 2020, 2021, and 2022 as part of Snowmass’21. More than three thousand scientists have taken part in the Snowmass'21 discussions and about 1400 people participated in the Seattle workshop in person and remotely. In general, the international community was very well represented and many scientists from Europe and Asia participated as organizers of sessions and events, conveners of topical groups, and by submitting numerous Letters of Interest (short communications) or White Papers (extended input documents). 

The key questions for the AF included:\\  
\emph{What is needed to advance the physics? \\
What is currently available (state of the art) around the world? \\
What new accelerator facilities could be available in the next decade (or next next decade)? \\ 
What R\&D would enable these future opportunities? \\
What are the time and cost scales of the R\&D and associated test facilities, as well as the time and cost scale of the facilities?\\} 

There were 7 AF topical groups led by internationally recognized researchers: AF1 "Beam Physics and Accelerator Education" - Mei Bai (SLAC), Zhirong Huang (SLAC), Steve Lund (MSU); AF2	"Accelerators for Neutrinos" -  John Galambos (ORNL), Bob Zwaska (FNAL), 	Gianluigi Arduini (CERN); AF3 "Accelerators for EW/Higgs" -  Angeles Faus-Golfe (IN2P3), Georg Hoffstaetter (Cornell), 	Qing Qin (ESRF), Frank Zimmermann (CERN); AF4 "Multi-TeV Colliders"  - Mark Palmer (BNL), Nadia Pastrone (INFN), Jingyu Tang (IHEP), Alexander Valishev (FNAL); AF5 "Accelerators for Physics Beyond Colliders and Rare Processes" - Mike Lamont (CERN), Richard Milner (MIT), Eric Prebys (UC Davis); AF6 "Advanced Accelerator Concepts" -  Ralph Assmann (DESY), Cameron Geddes (LBNL), Mark Hogan (SLAC), Pietro Musumeci (UCLA);  AF7 "Accelerator Technology - RF" Emilio Nanni (SLAC), Sergey Belomestnykh (FNAL), Hans Weise (DESY); "Magnets" -	Susana Izquierdo Bermudez (CERN),	Gianluca Sabbi (LBNL), Sasha Zlobin (FNAL); "Targets/Sources" - Charlotte Barbier (ORNL), Frederique Pellemoine (FNAL), Yin-E Sun (ANL). 

More than 300 Letters of Interest and 116 White Papers were submitted to the Snowmass'21 AF topical groups.  There were more than 30 topical workshops, 8 cross-Frontier {\it Agoras} (5 on various types of colliders: $e^+e^- / \gamma \gamma$, linear/circular, $\mu \mu$, $pp$, advanced concepts and 3 on experiments and accelerators for rare processes physics). Several special cross-Frontier groups were organized between  AF,EF,TF,IF, and NF such as {\it the eeCollider Forum, the Muon Collider Forum, the Implementation Task Force} (see below), the 2.4MW proton power upgrade design group at FNAL, etc.
All the AF LoIs, White Papers, links to all the workshops, topical group reports, the fora and the ITF report are available at the AF wiki page https://snowmass21.org/accelerator/start. 

Below we present summaries on all the AF topics in three major categories: future facilities, accelerator R\&D, and workforce education and training. Together with the above Executive Summary, they represent the Snowmass'21 Accelerator Frontier input to the Snowmass'21 final report and the 2023 P5 prioritization process. 


\section{Accelerator facilities for neutrino research (NF)}
\label{AF2summ}

The leading accelerator-based facilities for high energy neutrino research are {\it superbeams} based on a conventional beam dump technique: an intense high energy proton beam is directed onto a thick nuclear target producing mostly pions and kaons that are captured by focusing magnetic horns in order to obtain a well directed beam of same charge secondaries. High-energy neutrino beams are products of the decays of charged  pions  and  kaons  in  a long decay  channel \cite{shiltsev2020superbeams}. Superbeams sources operate with proton beam intensities close to the mechanical stability limit of the primary targets that is at present $O$(1 MW). The most powerful accelerators for neutrino research to date are the rapid cycling synchrotron facility J-PARC in Japan, that has reached 515 kW of 30 GeV proton beam power, and the Fermilab Main Injector delivering up to 893 kW of 120 GeV protons on target. These facilities support neutrino oscillation research programs at the SuperK experiment (295 km from J-PARC) and MINOS (810 km from Fermilab). 

\begin{figure}[htbp]
\centering
\includegraphics[width=0.75\linewidth]{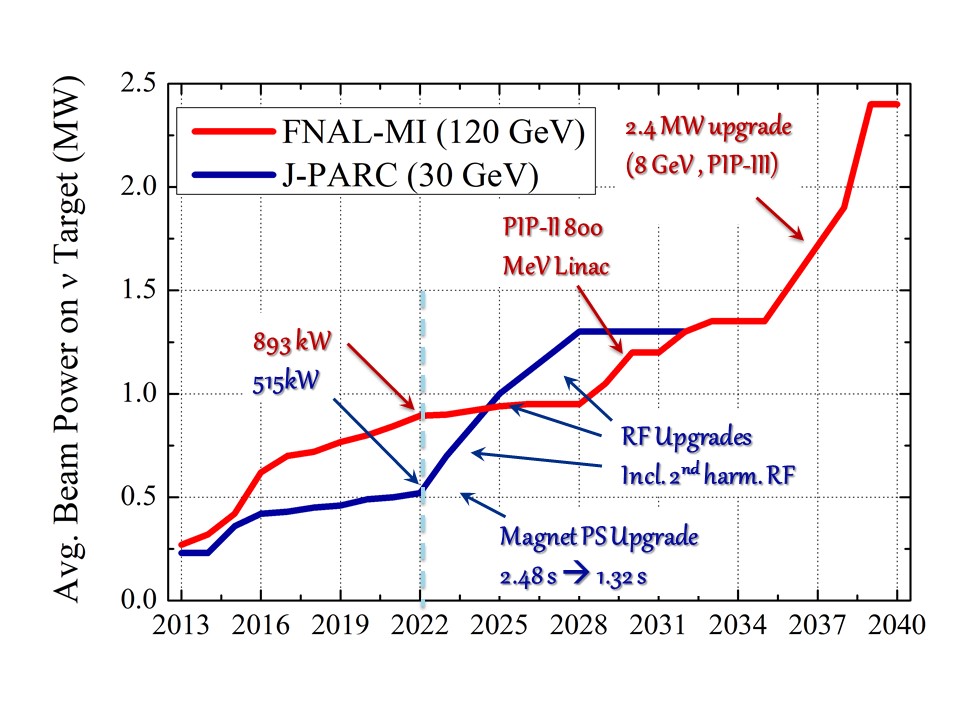}
\caption{Beam power progress and plans for J-PARC and the Fermilab Main Injector - two leading superbeam  facilities for neutrino research (adapted from \cite{shiltsev2020superbeams})} 
\label{Fig1}
\end{figure}

The needs of neutrino physics call for next generation, higher-power, megawatt and multi-MW-class superbeams facilities. Average proton beam power on the neutrino target scales with the beam energy $E_b$, number of particles per pulse $N_{ppp}$ and cycle time $T_{cycle}$ as $P_b = (E_b N_{ppp}) / T_{cycle}$. Corresponding upgrades of the RF system and magnet power supply ramping rate $1/T_{cycle}$  have been initiated at J-PARC, while Fermilab has started construction of an 800 MeV SRF $H^-$ PIP-II linac (Proton Improvement Plan-II) that will help to boost $N_{ppp}$ and the Main Injector beam power to above 1.2 MW, and is considering further facility upgrades to get to 2.4 MW - see Fig.\ref{Fig1}. 

Fermilab is now studying several accelerator approaches to further double that beam power to 2.4 MW to an enhanced DUNE experiment (as part of the LBFN/DUNE Phase II) that is presented as the highest priority for the neutrino program in the 2030s. Multiple options have been proposed including an additional high
duty-factor linear accelerator, and either a synchrotron or storage ring for accumulation of intense beam current required for the experiments.  Corresponding design studies are needed to identify the most cost- and power-efficient solutions.  

Besides the J-PARC and FNAL accelerator complex upgrades, an additional high-power, long-baseline neutrino beam has also been proposed to use a 5MW beam from an upgraded ESS proton accelerator delivered to megaton-scale detectors to measure CP violation. The ESS neutrino Super Beam proposal is instructive in demonstrating some of the challenges, and
perhaps ultimate limitations, of pushing to ever higher beam powers. The proton beam must be highly compressed so as to have a low duty factor – order of microsecond pulses, to be useful for experiments
and be properly focused by pulsed horns. Linac-based beams, with their rather low peak current and high duty factor, must employ a ring to accumulate and compress $H^-$ beams to be useful in this way.
The 2.5 GeV energy is on the low-end for high-power neutrino beams where pion production yield is marginally less efficient and particularly the proton flux on the target station is incredibly high, such
that the target station must be split into four separate targets and focusing horns. 

Other accelerator-based neutrino approaches are underway for other searches within neutrino physics. Several forms of “decay-at-rest” neutrino sources (often from high-power spallation neutron sources)
have detected coherent neutrino interactions and offer potential approaches to sterile neutrino oscillations and dark sector searches. Additional opportunities exist in further exploiting the timing and
kinematics of beams to gain additional experimental reach. Hadron colliders have been found to be significant sources of neutrinos and proposals are being developed to detect them at the LHC and future colliders. The neutrino factory concept (and its related proposals) remains an option for further precision in long- and short- baseline neutrino physics.

\subsection{AF2 ("Accelerators for Neutrinos"): main findings and recommendations}

The R\&D needed for many of these next generation accelerators for the Neutrino Frontier vary widely - see the AF2 summary report \cite{AF2summary}: 
\begin{itemize}
\item Handling multi-megawatt beams proposed for future neutrino beams or for proton drivers for future muon colliders demand extremely robust targets able to withstand enormous thermal and mechanical stresses, ideally over long periods and with minimum maintenance.
\item Superconducting proton linear accelerators have been a prominent tool to generate high power proton beams, but these beams must be compressed and formatted to short beam pulses that neutrino beam experiments require. Thus, development into higher-current linac $H^-$ beams, $H^-$ injection, and high-current accumulation are
essential.  Significant progress has been made in the design of alternative laser stripping schemes [SNS] but so far no operating high intensity machine is routinely using laser stripping for charge exchange injection and further research and development should be pursued.
\item 
High current broadband or highly tunable RF is needed for synchrotrons and accumulator rings. 
\item 
Instabilities and loss
control techniques are vital to all of the above machines as losses of $\ll$ 1\% can become prohibitive to
operations, such that the machine’s performance is limited more by efficiency than maximum current or energy. New concepts such as FFAs, integrable optics, space-charge compensation with electron lenses, and recirculating linear accelerators could provide advantages to future machines. 
\item Minimization of beam losses is critical to minimize damage to components and allow maintenance in case of failure. This requires the understanding of loss mechanisms and the implementation of these effects in simulation codes. The simulation effort should be accompanied by development of advanced diagnostics with high dynamic range to accurately characterize the beam properties to benchmark simulations and for protection and tuning purposes.
\item Present and future high intensity circular machines such as FCC-ee and CEPC demand H$^-$ low energy front-ends to maximize the beam brightness and minimize losses.
\item Given the high beam power, maximization of the efficiency of future accelerators is mandatory to guarantee the sustainability and affordability of the research conducted with these machines in view of reducing the carbon footprint and contain the operation costs, particularly in the present international context of increasing energy prices.
\end{itemize}

%
%
%

\section{Accelerator facilities  for rare processes physics (RPF)}
\label{AF5summ}

Many existing and planned high intensity accelerators can be effectively used for fixed target experiments complementary to high-energy-frontier colliders \cite{jaeckel2020quest}.  For example, a recent CERN  study of  the {\it Physics Beyond Colliders} (PBC) \cite{beacham2019physics} resulted in numerous accelerator-based proposals, ranging from a gamma-factory \cite{budker2022expanding} to explorations of the dark sector to precision measurements of either strongly interacting processes or light, feebly interacting particles at beam dump facilities, such as, e.g., the SHiP experiment \cite{ahdida2019sensitivity} at the SPS North Area \cite{ahdida2019sps}.  

A similar study has been undertaken in the US as part of the Snowmass process. It considers a variety of potentially available beams ranging from 800 MeV to 120 GeV protons \cite{pellico2022fnal, toups2022sbn, apyan2022darkquest} to multi-GeV electrons \cite{akesson2022current}. Modifications of existing accelerator complexes and future dedicated scientific infrastructure should be considered for the next two decades through projects complementary to main stream applications of existing facilities. The main areas of interest include: a) low energy hidden sector searches for the QCD axion, and sub-eV Axion/ALP (helioscopes like, e.g., BabyIAXO/IAXO, haloscopes using resonant cavities (ADMX) or other methods (MADMAX), light-shining-through-walls experiments (e.g. JURA, STAX); b) Light Dark Matter searches in the MeV – GeV mass range target a parameter space of the Hidden Sector of special relevance to open questions in cosmology (proton beam dump experiments, e.g., BDF/SHiP NA62, MiniBooNE, SeaQuest; electron beam dump experiments NA64, LDMX, BDX, etc or novel use of existing facilities such as LCLS-II, CEBAF, FAST/IOTA, TRIUMF/ARIEL); or long lived particles at colliders (LHC, SuperKEKB); c) 
precision measurements and rare decays that can probe higher masses than accessible with direct searches, via searches for the possible influence of the contribution of loop diagrams in a number of scenarios, e.g., 
ultra-rare or forbidden decays/reactions in the kaon sector - NA62, KOTO, KLEVER, lepton sector - TauFV, Mu3e, MEG,mu2e/mu2e-II, permanent EDM in protons/deuterons (CPEDM) or in strange/charmed baryons (LHC-FT), 
and the muon anomalous magnetic moment (g-2). 

\subsection{AF5 ("Accelerators for RPF"): main findings and recommendations}

Many of the above listed RPF initiatives can profit from ongoing advances in accelerator technology, e.g., high field superconducting magnets, superconducting RF, and are/could be considered at laboratories
that host this technology and associated technical infrastructure. There was also significant overlap of AF2 discussions with other working groups within the Accelerator Frontier, specifically AF2 (Accelerators for Neutrinos) and AF7-T (Accelerator Technology - Targets and Sources). 

Among the highest priorities in the AF5 "Accelerators for RPF" for the next decade are \cite{AF5summary}:
\begin{itemize}
\item Optimum exploitation of the PIP-II Linac at Fermilab: The PIP-II 800 MeV proton SRF linac is currently set to provide only 17 kW of beam power for LBNF/DUNE, about 1.2\% of the total potentially available beam of 1.6 MW. Individual experiments, like Mu2e-II, have come forward with fairly detailed proposals to use the 800 MeV beam directly from
the PIP-II linac[2] but as yet {\bf there is no coherent plan for beam distribution and beam usage}. Other muon experiments like $\mu \rightarrow  e\gamma$ and $\mu \rightarrow  3e$ could also be carried out at this facility, although those proposals are not yet as developed. Although it is not high energy physics, a surface muon program has been proposed as well. To increase the sensitivity of the search for muon to electron conversion, an FFA has been suggested to produce a pure muon beam with a narrower energy and time spread. 

\item A new 0.8-1 GeV proton bunch compressor ring (PAR) has been proposed to accumulate beam from the PIP-II linac and extract it in intense bunches at rates of 100-1000 Hz. Such a compressor could serve not only Mu2e-II but also drive a set of dark sector searches based on beam dumps. Because of the relatively developed state of construction and planning for the PIP-II project, it is of priority to invest proper {\bf effort into the design of the PAR and development of the experimental programs it can support.}

\item Beam dumps support a broad range of experiments, generally searching for dark sector or other rare particles. These generally don’t have demanding beam requirements beyond the total power and can use a wide range of beam energies of proton and electron beams, often almost parasitically. The competitiveness of any proposal in terms of physics reach and cost need to be carefully evaluated from a global perspective. 

\item Other opportunities and synergies: of particular interest is the Belle-II experiment at KEK, and storage rings for electric dipole moment measurements (e.g. electrostatic storage rings of the sort that have been proposed to measure the proton EDM). The BSM/RPF searches and activities are often synergistic with other R\&D areas. These include overlaps in target development with ongoing R\&D for neutrino or muon experiments, high field magnet development for “light through walls” experiments, and high-$Q$ RF development that in this case is primarily for axion searches.

\end{itemize}

\section{Colliders for Higgs/EW research (EF)} 


Charged particle colliders -- arguably the most complex and advanced scientific instruments --  have been at the forefront of scientific discoveries in high-energy and nuclear physics since the 1960s \cite{shiltsev2021modern}. There are five electron-positron colliders in operation at present: DA$\Phi$NE in Frascati, Italy, VEPP-4M and VEPP-2000 in Novosibirsk, Russia, BEPC-II at IHEP, Bejing, and SuperKEKB at KEK, Japan. The SuperKEKB is an asymmetric $e^+e^-$ B-factory with 4 and 7 GeV beam energies, respectively. Since the startup in 2018, 
SuperKEKB is pushing the frontiers of accelerator physics with 
crab-waist collisions, a world-record luminosity  (for any collider type) of 4.7$\cdot 10 ^{34}$ cm$^{-2}$s$^{-1}$,
and a vertical rms beam spot size of 300 nanometers. The future goal is 6$\cdot 10 ^{35}$ cm$^{-2}$s$^{-1}$ (a whopping 30-times over its predecessor KEKB (1999-2010)), and a beam spot size of 50 nm.
SuperKEKB is an important test-bed for FCC-ee and other future electron-positron colliders, and a key facility for training the next generation of accelerator physicists.


\begin{figure}[htbp]
\centering
\includegraphics[width=0.75\linewidth]{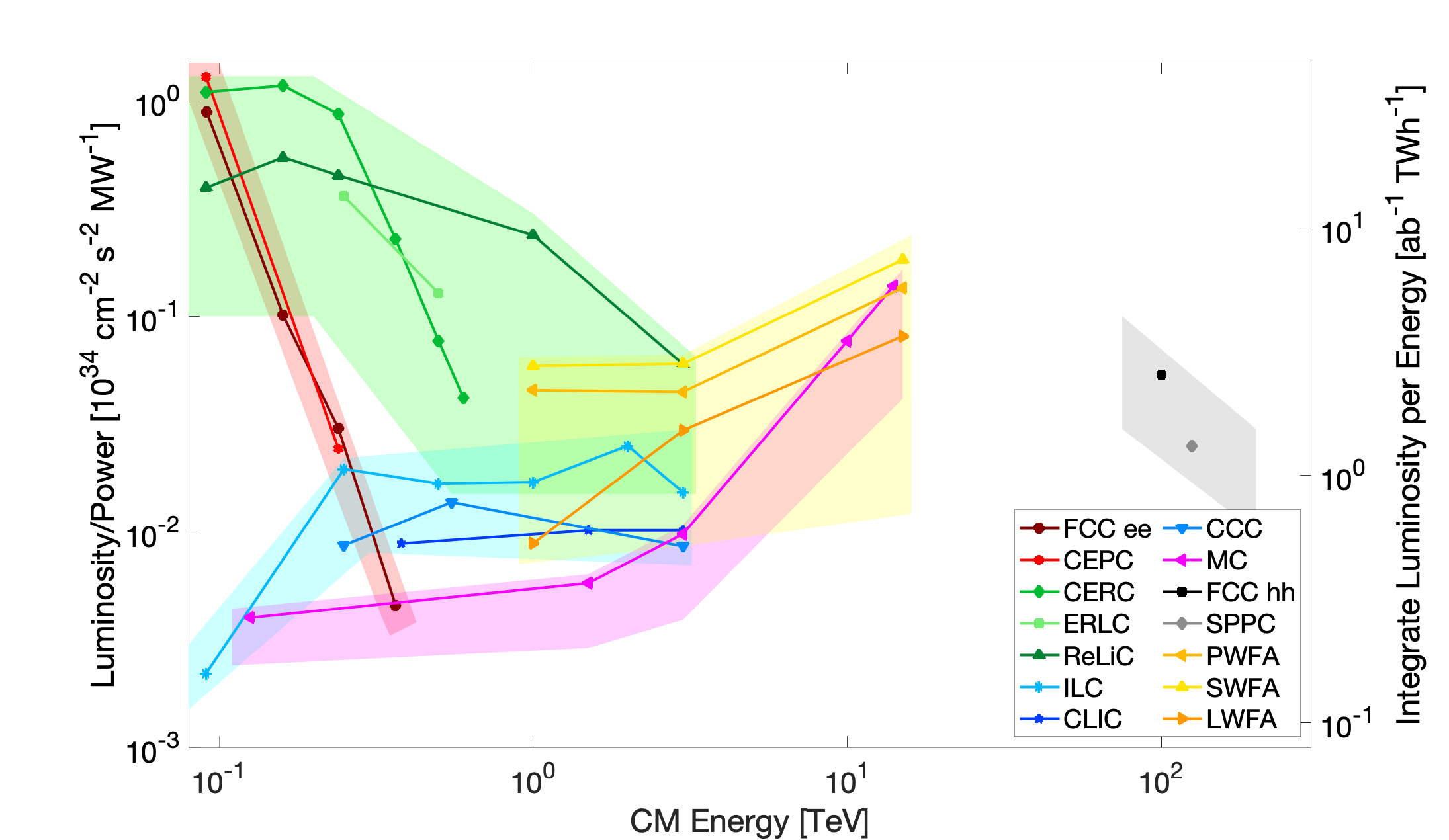}
\caption{Energy efficiency of present and future colliders. Luminosity per MW of the total facility power (left axis) and annual integrated luminosity per Terawatt-hour of electric power consumption (right axis) as a function of the centre-of-mass energy (from \cite{ITF}). The effective energy reach of hadron colliders (LHC, HE-LHC and FCC-hh) is approximately a factor of seven lower than that of a lepton collider operating at the same energy per beam. } 
\label{Fig2}
\end{figure}

\begin{figure}[htbp]
\centering
\includegraphics[width=0.75\linewidth]{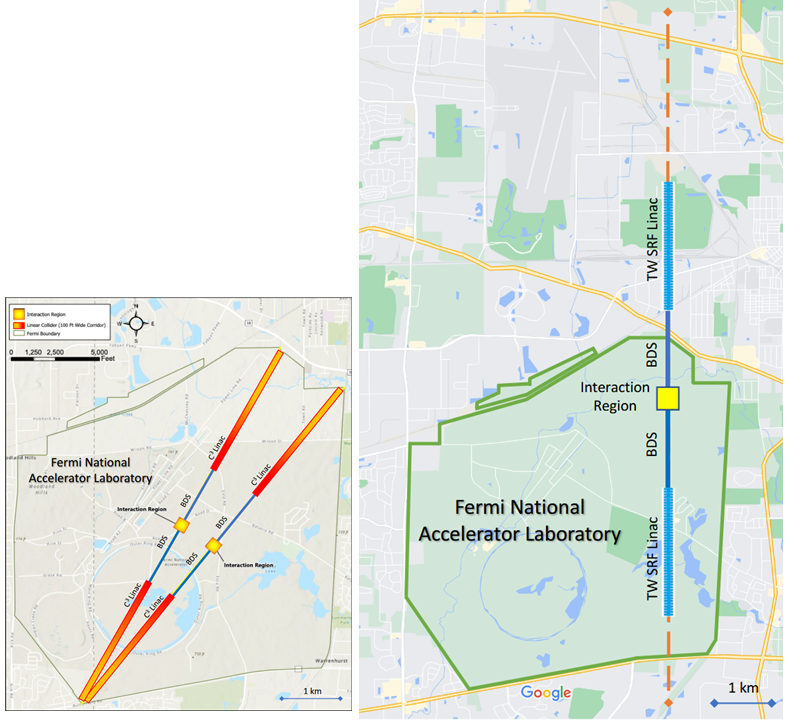}
\caption{Possible placements of future linear $e^+e^-$ Higgs/EW factory colliders $C^3$ and $HELEN$ on the Fermilab site map - both about the same length: (left) 250 GeV c.m.e. options with a 7-km footprint (from Ref.\cite{bhat2022future}), and (right) higher c.m. energy options (the orange dashed line indicates a 12-km stretch that might be made available for a future linear collider) \cite{HELEN}. } 
\label{Fig3}
\end{figure}

The world HEP community has called for a Higgs/EW factory as the next large accelerator project.  At present, a dozen Higgs/ElectroWeak factory proposals are under consideration, including $e^+e^-$ colliders such as the CEPC in China and FCC-ee at CERN, both on the order of 100 km circumference, that require $O$(100 MW) RF systems to sustain high luminosity \cite{lou2019circular}; or an 11 km long  CLIC (CERN) two-beam normal-conducting RF linear accelerator with an average 72 MV/m gradient \cite{stapnes2019compact}; or the 21 km long International Linear Collider (ILC) based on SRF 31.5 MV/m SRF linacs \cite{michizono2019international}. Two recently proposed Higgs factories can potentially be shorter than 7 km and fit the Fermilab site - $C^3$ (employs 5.7 GHz 70 MV/m cool copper RF at 77 K) \cite{bai2021c} and HELEN (1.3 GHz travelling wave SRF at 2K and 70 MV/m gradient) \cite{bhat2022future} - see Fig.\ref{Fig3}.  

Besides technical feasibility and affordable cost, the most critical requirements for future particle colliders include the center-of-mass energy reach, the required AC site power consumption (see Fig.\ref{Fig3}), and the required duration and scale of the R\&D effort to reach proper readiness -- see detailed discussion in Refs.\cite{shiltsev2021modern, ITF} and in Sec.\ref{ITF} below. Note that many of the future collider proposals have also been discussed in the course of the {\it 2020 European Particle Physics Strategy Update} \cite{abramowicz2019physics}. 

\subsection{Proposed Higgs/EW physics facilities}

Presented below are proposals for e$^+$e$^-$ Higgs factories - colliders that operate at 240/250 GeV and, alternatively, also colliders with a staged program, including a stage above the top quark threshold.  The latter colliders deliver superior physics performance for three reasons: (1)  The top quark is another important object that needs precise study. (2)  Above 350 GeV, the primary Higgs boson production mechanism changes from e$^+$e$^-$ $\rightarrow$ ZH to WW fusion production of the Higgs. This capability is essential to prove the influence of beyond Standard Model physics on the Higgs boson.
(3)  Some measurements require data at two different, sufficiently well separated c.m. energies.  The most important of these is the determination of the Higgs self-coupling from single Higgs production.

Most of the proposed Higgs/EW factories propose a significant run around the Z pole, opening a capacity for physics Beyond the Standard Model (BSM) through QCD and EW precision measurements. Analysis of the environmental footprint during construction and operation  indicates significant differences among the Higgs/EW factory proposals \cite{blondel2022footprint}. 

{\it International Linear Collider  (ILC).} The ILC is the most technologically mature of the proposed next-generation $e^+e^-$ colliders - see \cite{michizono2019international}. The initial phase of the project has a c.m.e. of 250 GeV for precision studies of the Higgs boson, a major goal in collider physics. The ILC design assumes a high degree of polarization in both $e^-$ and $e^+$ beams. As a linear machine it can operate at higher or lower energies. For example, the initial 21 km length can be extended to reach the threshold for the production of a top antitop quark pair ($t {\bar t}$) as well as for the associated production of a Higgs boson with a $t {\bar t}$ with c.m. energies up to 500 GeV with a length of 31 km. The site allows an upgrade to 1 TeV by extending the SRF linac lengths and, with improvements to the SRF cavities, c.m. energies of 3-4 TeV may be possible. 

At 250 GeV, the primary enabling technology is Superconducting RF (SRF) cavities operating at an average gradient of 31.5 MV/m. The European X-ray Free Electron laser (XFEL) in Hamburg Germany is a multi-billion dollar project that provides a 10\%-scale demonstration of the ILC acceleration systems with over 750 SRF cavities operating at an average of 23 MV/m and producing a 17.5 GeV electron beam for the XFEL. A Technical Design Report (TDR) for the ILC has been completed and the project is considered to be “shovel ready.” At this point, the R\&D focus is on reducing the cost of the cryomodules. The remaining technical challenges are: improvement of the positron source, achieving the nanometer-scale spot size and long-term stability and reproducibility at the interaction point (IP), and optimizing the damping ring injection and extraction systems.  

The ILC would be the first truly global accelerator project. An International Development Team (IDT) was established by the International Committee for Future Accelerators (ICFA) in August 2020 to pave a way for the preparatory phase of the ILC. Hosted by KEK, it is intended to support the Japanese HEP community effort to host the ILC as a global project. 
It coordinates a continued effort in accelerator R\&D for the engineering design. It also re-evaluates the ILC roadmap taking into account the global situation with an aim to trigger governmental discussion on the ILC as a global project that could lead to negotiations for the distribution of the cost and responsibilities for the construction and operation.

{\it Compact Linear Collider (CLIC).} The Compact Linear Collider (CLIC) is a multi-TeV high-luminosity linear $e^+e^-$ collider proposed by an international collaboration led by CERN - see \cite{stapnes2019compact}. The design is based on a staged approach that includes three c.m. energies of 380 GeV, 1.5 TeV and 3 TeV. In contrast to the ILC, CLIC uses a novel two-beam acceleration technique with normal-conducting accelerating structures operating in the range of 70 MV/m to 100 MV/m, and has only one beam polarized ($e^-$). A CDR was produced for CLIC in 2012 and, while the design is less mature than the ILC, the CLIC design parameters are well understood and have been reproduced in beam tests indicating that the CLIC performance goals are achievable.

The main risks and uncertainties for CLIC will be in scaling from the small-scale low-power demonstrations to the km-scale high-power two-beam deceleration systems. Efforts to reduce power consumption are ongoing and new estimates show a significant reduction related to improvements in the X-band RF technology and klystron design. Like the ILC, other R\&D includes improvement of the positron source,  achieving the nanometer-scale spot size and long-term stability and reproducibility at the interaction point (IP), and optimizing the damping ring injection and extraction systems.  In general, the spot sizes and stability requirements are tighter for CLIC than for the ILC while the positron system requirements may be easier due to a lower number of $e^+$/sec and a higher macro-pulse repetition rate that eases requirements on a rotating target.

{\it Future Circular e$^+$e$^-$ Collider (FCC-ee).}
 The Future Circular Collider (FCC) is a proposed international collider complex located near Geneva Switzerland. It is based on the same successful staging strategy used for the Large Electron-Positron collider (LEP) and the Large Hadron Collider (LHC). The approximately 100 km tunnel would initially house the FCC-ee $e^+e^-$ collider that would offer a broad physics reach operating at four different c.m. energies: the $Z$ pole, $WW$ threshold, the $ZH$ production peak and the $t {\bar t}$ threshold.  The tunnel could eventually house the FCC-hh, a 100 TeV proton-proton collider - see \cite{benedikt2019physics}.  A Conceptual Design Report (CDR) for the FCC complex was completed and published in 2019 \cite{fcc}; \cite{fcchh}; \cite{fccee} and a Siting and Feasibility Study for the FCC-ee \cite{council-fcc-1}; \cite{council-fcc-2} will be completed in 2025, making it one of the more advanced proposals.  

The main technologies for the FCC-ee are well-developed. The technology R\&D is focused on incremental improvements aimed mainly at further optimizing electrical efficiency, obtaining the required diagnostic precision, and achieving the target performance in terms of beam current and luminosity.  Optimization is also desired to improve the performance of the positron source, SRF and vacuum systems \cite{FCCsnowmass}. The FCC Feasibility Study, launched by the CERN Council in 2021, will address numerous key feasibility aspects, including tunnel construction, financing, sustainability and environmental impact. The result of the Feasibility Study will be an important input to the next European Strategy Update expected in 2026/27. The FCC technical schedule foresees the start of tunnel construction in the early 2030s, the first  $e^+e^-$ collisions at FCC-ee in the mid or late 2040s, and the first FCC-hh hadron collisions around the year 2070.  

{\it Circular Electron Positron Collider (CEPC).}
 Proposed by Chinese scientists in 2012, the CEPC is an international scientific project hosted by China to build a 240 GeV circular $e^+e^-$ collider in an approximately 100 km tunnel - see \cite{lou2019circular}. Similar to the FCC, the tunnel for the CEPC could eventually be used for a Super Proton Proton Collider (SPPC). A CDR for the CEPC was released in November 2018. The TDR is planned for completion at the end of 2022 followed by work on an Engineering Design Report (EDR) that will look at the detailed engineering design of components, site selection and preparations for industrialization.

Like the FCC-ee, the technological basis for the design is well-understood and the R\&D focus is now on improving performance of the RF klystrons, SRF cavities, high precision magnets, and vacuum systems.  The proposed schedule calls for first collisions in the mid-2030’s.

{\it Cool Copper Collider (C$^3$).}
 The C$^3$ is a linear collider concept based on recent innovations in the technology of cold copper cavities providing power to each individual cell that allows for increased accelerator performance and better optimization - see \cite{bai2021c}. Operation at liquid nitrogen temperature substantially increases the RF efficiency and gradient of normal-conducting copper cavities. For a 250 GeV c.m. energy the accelerator is 8 km and, with additional innovation in cost-efficient RF power sources, 550 GeV could be reached using the same footprint. A GeV-scale demonstration facility is proposed to provide input for a TDR. It would include three cryomodules operating at 70, 120 and 170 MV/m to test the RF design. A gradient of 155 MV/m would be needed for the 550 GeV upgrade. Other collider subsystems can be based on those developed for CLIC and ILC.  Areas of technical focus are development of an ultra-low emittance polarized electron gun that would benefit all linear collider designs, optimization of the RF structures and reducing the cost of the RF sources.
 
 {\it HELEN.} The Higgs-Energy LEptoN (HELEN) linear $e^+e^-$ collider proposal is based on advanced travelling wave (TW) 1.3 GHz superconducting radio frequency technology \cite{HELEN}. Relative to the ILC, the proposed collider offers cost and AC power savings, a smaller footprint  $\sim$7.5 km (vs 20.5 km), and could be built at Fermilab with an Interaction Region within the FNAL site boundaries. After the initial physics run at 250 GeV c.m.e., the collider could be upgraded either to higher luminosity or to higher (up to 500 GeV) energies. The proponents' cost estimate for HELEN is about 15\% less than the ILC-250 construction cost. If the ILC can not be realized in Japan in a timely fashion, the HELEN collider would be a viable option to build a Higgs factory in the U.S. All collider subsystems with the exception of the TW SRF can be copied from the ILC design. Corresponding R\&D is needed to reduce technical cost risks, demonstrate the TW SRF cryomodules with 70 MV/m gradients and efficient $e^+$ generation - as for the ILC. 
 
{\it Energy Recovery and Recirculating Linacs (ERLs and RLAs).}
 Linac-based and circular colliders with energy recovery are an alternative approach to high energy electron-positron colliders with the aim to significantly reduce beam energy losses and consequently, power consumption. There are two proposed configurations. A circular $e^+e^-$ collider with two 100 km storage rings using Energy-Recovery Linacs (CERC) \cite{litvinenko2020high}
 \cite{litvinenko2022cerc} or two large linear colliders with damping rings, a Linear Energy Recovery Linac Collider (ReLiC) \cite{litvinenko2022relic} and an ERL-based linear $e^+e^-$ collider ERLC \cite{telnov2021high}. Starting as a Higgs factory they have the capability of achieving c.m. energies up to 600 GeV. The energy as well as the particles are recycled in this scheme and make fully polarized electron and positron beams possible. A large fraction of the energy of the used beams is recovered by decelerating them. The beams are then reinjected into a damping ring where they are cooled and reused. Beam that is lost during the recovery process is replaced via a linear injector into the damping rings. 
 
 ERL's are currently in a process of  technological development.  A number of compact ERL facilities and demonstrators have been constructed.  The highest circulating Continuous Wave (CW) power was achieved in the IR FEL ERL at Jefferson Laboratory that operated with 8.5 mA at 150 MeV.  A global development program for the ERL technology is discussed in \cite{CERN2022RnD}.
 
{\it X-ray Free Electron Laser Compton Collider (XCC).}
 The XCC concept combines the cold copper distributed coupling technology of C$^3$ with X-ray FELs \cite{barklow2022xcc} to create a $\gamma$-$\gamma$ collider. The linac accelerates electron bunches with a gradient of 70 MV/m until 31 GeV is reached, at which point alternate bunches are diverted to an X-ray FEL to produce circular polarized 1 keV X-rays using a helical undulator. The electron bunches remaining in the linac are accelerated to 62.8 GeV through a final focus system to the $e^-e^-$ interaction points (IP). The 62.8 GeV electrons then collide with the focused X-ray laser light from the opposite X-ray FEL, producing 62.5 GeV photons that are then collided at 125 GeV c.m.e. in the primary IP. The number of Higgs' produced in such a machine would be comparable to the ILC, but backgrounds need to be studied in more detail. 
 
 Relative to some other $e^+e^-$ Higgs factory proposals the XCC requires two additional beamlines and collision points and requires significant improvement to X-ray FEL technology. However, if a high brightness polarized RF gun can be developed, damping rings would not be required and the beam energy is half of that required for $e^+e^-$ Higgs factories, raising the possibility of significant reduction in cost, a common obstacle for large-scale facilities. Gun development is challenging, but will benefit from experience with the LCLS-II-HE SRF gun when it turns on. 
 
\subsection{AF3 ("Higgs/EW Factories"): main findings and recommendations}
\label{AF3summ}

Comparisons of the main parameters of the proposed Higgs/EW factories were made for the linear colliders, the circular colliders and the $\gamma\gamma$ colliders. Critical R$\&$D items were identified for each proposed collider, and consolidated R$\&$D efforts that would jointly benefit several of the projects have been sketched out. Most of the proposed Higgs and Electroweak Factories can not be considered in isolation from other accelerators that are being  proposed for higher energies. Higher energy machines would either greatly benefit from the infrastructure and accelerator developments described here, or would be a direct upgrade or extension of the Higgs and Electroweak Factory. This provides a natural connection to the accelerator topical group AF4 "Accelerators for Energy Frontier", see below in Sec.\ref{AF4summ}.

Major conclusions of the AF3 group are \cite{AF3summary}:
\begin{itemize}
    \item Of all the Higgs/EW factory proposals considered, only a small number of them are ready or close to a construction phase - namely, the ILC, CLIC, FCC-ee and CEPC. 
    \item Most of the other proposals are very high-level and in the conceptional design stage. These proposals should focus on the main R$\&$D tasks to move forward to a TDR. A detailed estimation of the Technical Readiness Level (TRL) risk factor, technology validation, cost reduction impact, performance achievability and timescale to reduce the TRL are presented in detail in the Collider Implementation Task Force report \cite{ITF} (see also, below in Sec. \ref{ITF}).
    \item The Higgs/EW factories centered on 240-250 GeV energy are not very challenging from the point of view of energy, but they are high-precision machines and luminosity will be the main figure of merit. Therefore, the focus of the R$\&$D effort should be on how to make these colliders more compact, reducing the cost, and to assure the highest possible integrated luminosity within reasonable site power limits and duration of operation.  
\end{itemize}

The joint technology research topics have been identified that will be most beneficial for all Higgs/EW facotory colliders:
\begin{itemize}
\item Energy: a) SRF: TW structures and Nb$_{3}$Sn (70 MV/m HELEN). Special concern has to be paid to the SRF for ERLs; b) NCRF: Cryo-cooled Copper structures (120 MV/m C$^3$), HTS coatings; c) cryogenics: mass industrial production, transport issues, gas-pressure regulations, more efficient gas coolers.

\item Luminosity: a) positioning, monitoring, alignment and stabilisation: global strategies, instrumented girders, radiation-hard ground motion sensors; b) $e^+$ production optimization: flux concentrators, pulsed solenoid, capture linacs, targetry issues; c) nanobeam colliding techniques: concepts and feedback; d) damping rings and booster: low-emittance  and 4th generation lattices for colliders; e) magnets: Interaction Region final focus systems and injection/extraction devices.

\item Sustainability: a) Energy consumption, efficiency, carbon footprint; b) high-efficiency RF power sources: klystrons, solid state amplifiers and IOTs; c) permanent magnets.

\item Other: a) manufacturing techniques including additive, cost reduction and mass production; b) high power beam dumps (multi-MW); c) 
machine protection and collimation; d) polarized beams and polarimetry; e) beam instrumentation; f) robotics and automatization. 
\end{itemize}

The main technical topics to be worked on for individual proposals are:
\begin{itemize}
    \item ILC: 
     polarized e$^{+}$ production,  final focus system (FFS), tunability and long term stability,  FFS doublet vibration issues, Injection/extraction devices.
    For upgrades: SRF with higher Q, higher gradient, Traveling Wave SRF cavities, and  Nb$_{3}$Sn.

    \item CLIC: mechanical vibration mitigation, cost efficiency, X-band components and RF sources. Cavities with partial HTS coating.
    
    \item C$^{3}$: RF optimized structure including cost reduction and industrialisation, cryomodule R$\&$D, and an ultra-low emittance e$^{-}$ source. Overall accelerator layout.
    
    \item HELEN: SRF with higher Q, higher gradient, Traveling Wave SRF cavities, and  Nb$_{3}$Sn. Cryogenic optimization. Overall accelerator layout. 
    
    \item ReLiC: SRF with higher Q, higher gradient, Traveling Wave SRF cavities, and  Nb$_{3}$Sn. Cryogenics optimization. Test of high current, low loss energy recovery.  Self-consistent and coherent parameter table and overall accelerator layout.
    
    \item ERLC: Dual-axis SRF cavities, higher Q, higher gradient. Cryogenics optimization. Test of high-current, low loss energy recovery.  Self-consistent and coherent parameter table and overall accelerator layout.
    
    \item XCC: X-ray transport and focusing. Interaction-region layout with Compton collision point. FEL design. Overall accelerator layout and integration issues.
    
    
    \item FCC-ee: Single- and few- cell 400 MHz Nb/Cu cavities with high $Q_0$, two- and five-cell 800 MHz bulk Nb cavities; cryomodule design; efficient RF power sources at 400 and 800 MHz; efficient cryogenic system; ``low-field" HTS magnet systems for the FCC-ee collider final focus, collider arcs, and for the positron source; magnet system for the fast ramping full-energy booster. High-field magnet systems based on Nb$_3$Sn and/or HTS in preparation of FCC-hh.
    
    \item CEPC: two-cell 650 MHz bulk Nb cavities; efficient RF power sources at 650 MHz and 1.3 GHz; cryogenic system; booster magnets. High-field magnet systems using iron based HTS for SPPC.
    
    
        
    \item CERC: SRF with higher Q, ultra-small emittance preservation, damping rings with very flat beams and large energy acceptance, use of small gap magnets for power and cost reduction. High repetition rate extraction and injection kickers. Self-consistent and coherent parameter table and overall accelerator layout.
    
    
\end{itemize}

\section{Collider facilities for the energy frontier research with $O$(10 TeV/parton) c.m.e. (EF)} 

Currently, there are two operational high-energy hadron colliders - RHIC at BNL and LHC at CERN. The Large Hadron Collider now represents the "accelerator frontier" with its 6.8 TeV energy per beam, 2.1$\cdot 10 ^{34}$ cm$^{-2}$s$^{-1}$ luminosity and some 1 TWh of annual total site electric energy consumption. Since the start of operation in 2009, the LHC has delivered 190 fb$^{-1}$/IP in $pp$ collisions -- mostly over the past few years at 13 TeV c.m.e., exceeding its design luminosity goal by a factor of two. The High-Luminosity LHC upgrade will be completed by 2028 with the goal of reaching 250 fb$^{-1}$/yr at 14 TeV c.m.e. via doubling the beam current, lower beta-function at the IPs with new Nb$_3$Sn SC IR magnets, and using {\it beam crabbing} and {\it luminosity leveling} techniques \cite{bruning2019high}.  The upgrade will be followed by a decade of operation to obtain a total integrated luminosity of 3-4 ab$^{-1}$. 

Two colliders are under construction - and both will serve the nuclear physics research community - the NICA pp/ion-ion collider in Russia with $\sqrt{s}$=4-11 GeV with expected first collisions ca 2024 and the Electron-Ion Collider (EIC) in the US (BNL) that will collide 10-18 GeV electrons with ions/protons up to 275 GeV energies. The later will be an excellent test bed for HEP colliders. Many topics, including polarization, ERLs, cooling, etc. will be developed for this machine and will benefit future high energy colliders. 

There are also about two dozen proposed energy frontier colliders that go beyond LHC in their discovery potential. Among them are the 3 TeV CLIC option (100 MV/m accelerating gradient, 50 km long), two roughly 100~km circumference $pp$ colliders, the SPPC in China (75-125 TeV c.m.e., based on 12-20 T SC magnets) and FCC-hh at CERN (100 TeV, 16 T) \cite{benedikt2019physics}, and a {potentially}  more economical 10--14 TeV c.m.e. $\mu^+\mu^-$ collider (10--14~km circumference, 12-16 T magnets) \cite{long2021muon}. {R\&D is in progress on other concepts such as wakefield based $e^+e^-$ or $\gamma \gamma$ systems that may present additional future options.} 

\subsection{Proposed energy frontier colliders}

{\it Future Circular Collider - hadron/hadron (FCC-hh).}
Proposed as a second phase of the FCC program after FCC-ee, the FCC-hh is a proton-proton collider aimed at increasing the physics reach by an order of magnitude beyond the LHC \cite{benedikt2022future}. Following the same strategy used for LEP and LHC, the accelerators will utilize the same 100 km-scale tunnel. It is one of the relatively mature proposals, and a conceptual design report was published in 2019 \cite{fcchh}. One of the major enabling technologies are the 16 to 17 T superconducting magnets required to reach the target energy of 100 TeV. This field level does not currently exist, but programs in the US and EU/CERN are actively pursuing the R\&D. Experience from the construction of high gradient quadrupoles for the LHC high luminosity (HL-LHC) upgrade will serve as a launching point for further development.

Attaining the desired luminosity of 30$\cdot$10$^{34}$ cm$^{-2}$s$^{-1}$ will be challenging. However, the HL-LHC will be an opportunity to gain considerable experience. Other challenges are related to the size and number of components, and increased number of injector rings, adding considerably to the overall complexity. Crab cavities, necessary for compensating the crossing angle at the interaction points, are still an untested system in hadron colliders, but the HL-LHC will provide an opportunity to develop this technology.

{\it Super Proton-Proton Collider (SPPC).}
The SPPC is proposed as the second phase of the CEPC-SPPC, sharing the same tunnel in a scheme parallel to the FCC-ee/hh \cite{tang2022snowmass}. It is planned to operate at a c.m. energy of up to 125 TeV in the final stage using 20 T magnets with an intermediate stage at 75 TeV using 12 T magnets. 

High-field magnets are the key enabling technology for future hadron colliders. A unique feature of the magnet technology proposed for the SPPC is the use of iron-based high temperature superconductor (IBS). This material is still in the early R\&D stage, but if successful, has the potential for significant cost savings over currently available HTS materials. A few programs in the US and EU are also engaged in exploring the potential of IBS. The project is currently in the pre-CDR phase. In addition to the high field magnets, synchrotron radiation power, luminosity and site power are a consideration in the overall design \cite{tang2022design}.

{\it Collider-In-the-Sea.}
The Collider-in-the-Sea is a 500 TeV c.m.e., 2100 km hadron-hadron collider to be located underwater in the Gulf of Mexico using low cost superferric SC magnets.

\begin{figure}[htbp]
\centering
\includegraphics[width=0.75\linewidth]{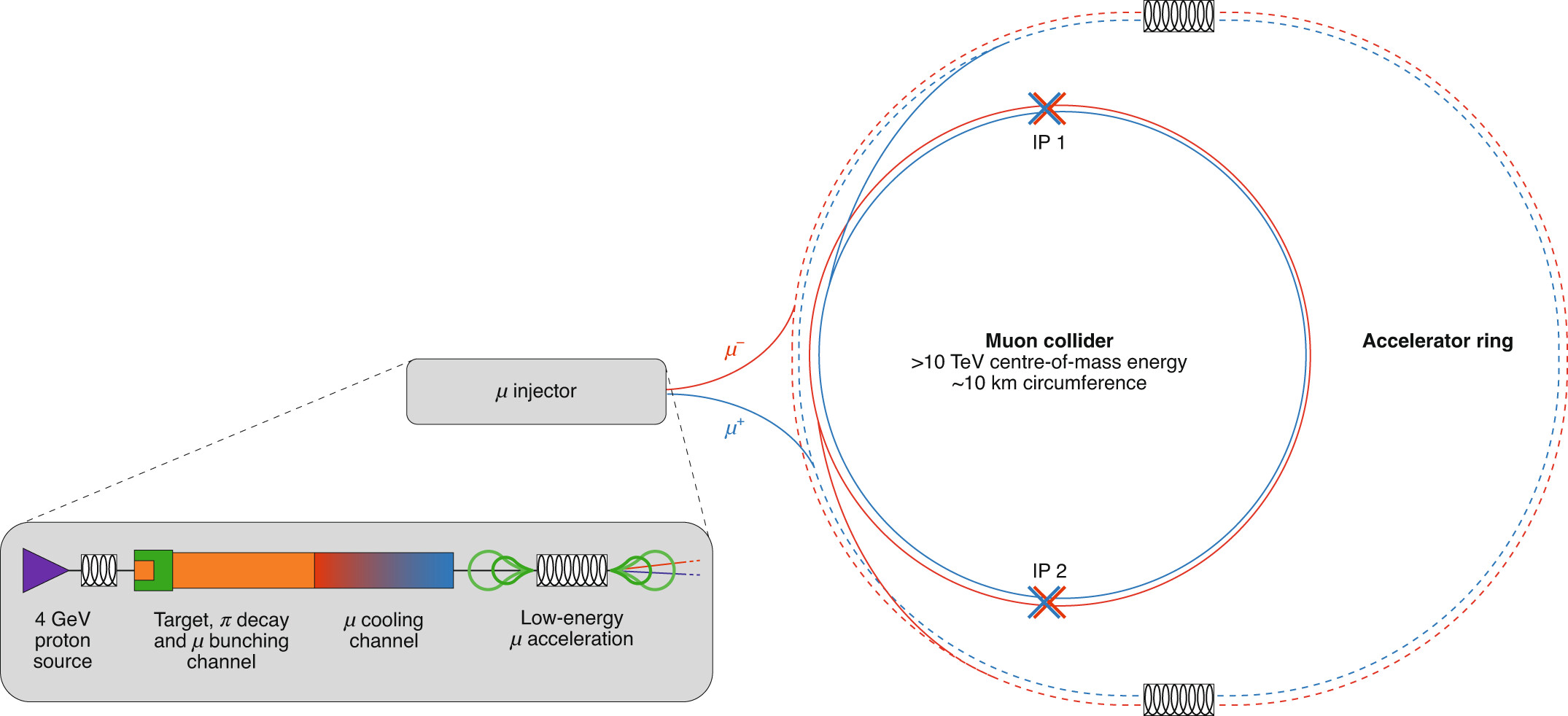}
\caption{A conceptual scheme for the muon collider (from Ref.\cite{long2021muon})}. 
\label{Fig4}
\end{figure}

{\it Muon Colliders.} The muon collider has great potential to extend the energy of lepton colliders by taking advantage of the strong suppression of synchrotron radiation from muons relative to electrons, though the finite lifetime of the muon is a critical issue. This allows for  efficient acceleration in rings and a more compact RF system - see Fig.\ref{Fig4} - as well as a better defined collision energy. For example, the energy consumption of a 10 TeV c.m. energy muon collider is estimated to be lower than CLIC at 3 TeV - see \cite{long2021muon}. As a ring, a muon collider may be able to provide luminosity to two detectors. 
 
 There are a number of technological challenges that need to be addressed to be able to take advantage of the large energy and luminosity reach \cite{CERN2022RnD}. To this end, an International Muon Collider Collaboration (IMCC) was formed based on a recommendation in the update of the European Strategy for Particle Physics \cite{ESPPu}. The initial focus is on 10 TeV with an integrated luminosity of 10 ab$^{-1}$.

The main critical enabling technologies are 6D cooling of the muon beams, development of high field solenoids and accelerator magnets, very fast ramping magnets, high power targets, a proton driver, and potential strong beam instabilities. There has been recent progress in most of these areas demonstrating feasibility, but in the near term, a robust R\&D program will be needed to bring the technology to the same level as linear or circular lepton colliders \cite{CERN2022RnD}.  

The scale of the R\&D program can be expected to be a significant fraction of the final facility cost, much as for the linear collider program where a large fraction of a B\$ has been invested in dedicated international test facilities beyond the Stanford Linear Collider.  Some of the R\&D could be done via an initial stage of 3 TeV at approximately half the cost of the 10 TeV version. Assuming sufficient funding to support a technically limited program and successful development of key technologies, it may be feasible to start colliding beams in the mid 2040’s.

{\it Fermilab Site Fillers.} Domestically, the US is fully engaged in the Long Baseline Neutrino Facility (LBNF) and the Deep Underground Neutrino Experiment (DUNE). However, given the long time-frame needed to plan and build large accelerator facilities, preparation for major new projects needs to start soon. Future energy frontier collider concepts that might feasibly be built on the Fermilab site include a 16 km circumference circular proton-proton collider with (24 – 27 TeV) c.m.e. and a staged muon collider from a Higgs factory at 125 GeV up to 8 – 10 TeV.

Each of the above options of hadron-hadron and muon colliders will require varying levels of R\&D to produce a CDR by the time of the next Snowmass, ca. 2030. To accomplish this, it is proposed that the U.S. establish an integrated future colliders R\&D program in the DOE Office of High Energy Physics (OHEP) to carry out feasibility studies and collaboratively engage in projects proposed abroad - see below Sec. \ref{IFCsumm}.

{\it Colliders based on wake-field acceleration concepts.}
These concepts are further discussed in Section 7.9.4 as an R\&D topic because self-consistent parameter sets are not yet available -- see Appendix 7.6 of Ref. \cite{ITF} for the latest suggested parameters.

Electric fields due to charge separation in plasma can sustain $> $GV/m gradients and have enormous promise for accelerator technology - see, e.g., \cite{ferrario2021advanced}. To date, several experiments demonstrated $O$(1-10 GeV) acceleration over 0.1-10 m long plasma channels. In principle, plasma wakefield (PWFA) linear accelerators can be employed in high energy $e^+e^-$ and $\gamma \gamma$ colliders, but a significant long-term R\&D effort is required to address many critical issues, such as acceleration of positrons, beamstrahlung, staging of multiple plasma acceleration cells, power efficiency, emittance control, jitter and scattering in plasma, etc \cite{shiltsev2021modern}, \cite{CERN2022RnD}. 

Structure wakefield acceleration (SWFA) has an advantage for application to linear colliders because the structures naturally accelerate electrons and positrons and are expected to be less challenging in preserving the beam emittance than plasma accelerators. Recent progress in development has increased the gradient to 300 MV/m and wakefield power generation to 500 MW \cite{jing2022continuous}.

\subsection{ AF4 ("Multi-TeV Colliders"): main findings and recommendations} 
\label{AF4summ}

The ongoing Snowmass process has emphasized the emerging 
need to explore the Energy Frontier with particle collisions where the constituent center-of-mass energy is $E_{cm}$ (c.m.e.) $\ge$ 10 TeV.  Thus, the principal focus of the "Multi-TeV Colliders" group (AF4) has been a review of the machine options for hadron and lepton colliders that may provide a path to this threshold \cite{AF4summary}. A consensus is emerging that the earliest timescale for making a construction decision for such a discovery machine will be
sometime in the next decade. Thus, the evaluations were focused heavily on the maturity of the various concepts and the type of support that will be required to provide the HEP community with the design input required for a machine decision on that timescale or at the earliest date. Moreover there is the need to conduct the R\&D and the integrated design work required for an equitable comparison of collider options during the course of the high luminosity LHC research program. 

In addition to the frontier discovery machine concepts, various submissions were received describing lepton colliders that could operate in the 1-few TeV range, mostly as possible extended options of more
mature Higgs factory designs. Furthermore, interest remains in the
possibility of alternative paths exploring the TeV-scale including lepton-ion colliders and $\gamma \gamma$ colliders, with the latter perhaps driven by plasma or structure wakefields.

The AF4 evaluation of the various concepts - based on the current design maturity and R\&D maturity - is as follows \cite{AF4summary}:
\begin{itemize}
    \item At present only a few design concepts have achieved a level of design maturity to have reliable performance evaluations based on prior R\&D and design efforts. They are the linear $e^+e^-$ collider ILC (1 TeV), CLIC (reaching 3 TeV c.m.e.), FCCeh, based on LHeC design and FCChh. Critical project risks have been identified
and subsystem focus R\&D is underway where necessary. 
\item
Most of the emerging accelerator complexes still
require significant basic R\&D and design effort to reach a maturity level to be properly evaluated. Those could mainly profit from a directed R\&D plan and a set of test facilities to demonstrate a broad range of
technology concepts that are required. Dedicated demonstrators are generally necessary before a “reference” design can be completed. To this category it’s easy to assign the SPPC hadron collider, muon colliders reaching c.m.e. 10 TeV and the newly proposed $e^+e^-$/$\gamma \gamma$ colliders based on WFA. Moreover other lepton colliders with c.m.e. $\leq$5 TeV (multi-TeV ILC, ReLIC, CCC) still require further studies, as well as a Muon Ion Collider. 

\item It is clear that the concept of a Collider-in-Sea is not ready to be considered since a proof-of-principle R\&D is required and the design is at a low level of maturity.

\end{itemize}

It should be noted that the concepts in the second (middle) category have the potential to achieve sufficient maturity within the next decade for evaluation by the HEP community. It is important to note that the necessary technical maturity for these concepts, and hence the ability to evaluate both the overall physics performance as
well as cost scale, cannot be delivered without dedicated collider R\&D research investment over the next several years, e.g., as in the proposed integrated collider R\&D program, see Sec.\ref{IFCsumm}.

\section{Snowmass'21 future colliders fora (AF-TF-EF)}
\label{Fora}

\subsection{$e^+e^-$Collider Forum: main findings and recommendations}
\label{eeForumsumm}
The Snowmass'21 $e^+e^-$ Collider Forum was organized by the Energy, Theory and Accelerator Frontiers and charged "...to promote dialogue and discuss differences and complementarities between the various $e^+e^-$ collider concepts, either linear or circular, from the accelerator, detector and physics perspectives, in harmony with the rest of the wider international community". The Forum’s activities - led by Maria Chamizo Llatas, Sridhara Dasu, Ulrich Heintz, Emilio Nanni, John Power, and Stephen Wagner - were carried out in close collaboration with the topical group conveners of the relevant frontiers. The Forum discussions covered a broad range of future electron-positron colliders from $e^+e^-$ Higgs Factories - linear and circular - capable of providing a rich scientific program with sub-percent Higgs boson coupling measurements, to potential discovery machines for the next New Physics scales at $O$(10 TeV) c.m.e..

Key findings and recommendation of the $e^+e^-$Collider Forum report \cite{eeFORUMsummary}, related to accelerators, include: 
\begin{itemize}
\item Higgs factories: primary consideration for the delivery of physics results is the start time of the physics program. Given the maturity of the technology, the ILC and the large storage rings hold the advantage of a possible early start of the program.  The ILC and CEPC both discuss possible starts late in the 2030s.  The FCC-ee would follow the HL-LHC program and start in the mid to late-2040s.  C$^3$ discusses a start similar to that of ILC assuming the completion of a technology demonstrator.  There is no published timeline for HELEN at this time.
\item FCC-ee and CEPC have signiﬁcant luminosity advantage and are able to complete the required runs at various luminosities faster but their larger civil engineering work requires signiﬁcantly more time and cost. An early start of the civil engineering construction of a circular machine is therefore key to timely realization of physics.
\item The ILC, HELEN and C$^3$ have cost, higher energy-reach, and polarization advantages but with lower luminosity, needing signiﬁcantly longer running time to achieve the same level of precision for measurements compared to circular machines.  From a potential siting point of view all but the C$^3$ and HELEN machines require greenﬁeld sites.
\item Given the strong motivation and existence of proven technology to build an $e^+e^{-}$ Higgs Factory in the next decade, the US should participate in the construction of any facility that has ﬁrm commitment to go forward.
\item Development of a $O$(10)-TeV scale $e^+e^-$ machine based on wakeﬁeld acceleration with suﬃcient luminosity capability for $O$(10) ab$^-1$ and energy-recovery technologies for improved power-to-luminosity costs, requires continued R\&D investment. Corresponding accelerator R\&D should focus on development of self-consistent machine design parameters and feasibility of attainment of collider speciﬁcations for the energy frontier.

\end{itemize}

\subsection{ Muon Collider Forum: main findings and recommendations}
\label{MCForumsumm}
There has been a recent explosion of interest in muon colliders, as evident from a ten-fold increase in the number of related publications submitted to arXiv in the last couple of years. The topic generated a lot of excitement in Snowmass meetings and continues to attract a large number of supporters, including many from the early career community. In light of this very strong interest within the US particle physics community, Snowmass Energy, Theory and Accelerator Frontiers created a cross-frontier Muon Collider Forum in November of 2020. The Forum conveners - Kevin Black, Sergo Jindariani, Derun Li, Fabio Maltoni, Patrick Meade, and Diktys Stratakis (two from each of AF, TF, and EF) - invited many experts to give their perspective and to educate the broader community about physics potential and technical feasibility of muon colliders. It facilitated a strong bond and exchange of new ideas between the particle physics community and accelerator
experts. Synergies with the neutrino and intensity frontiers, as well as overlaps with nuclear science and industry applications were also extensively discussed. Finally, the Forum served as an interface between the US community and the International Muon Collider Collaboration (IMCC) hosted by CERN. 

Key findings and recommendation of the Muon Collider Forum report \cite{mumuFORUMsummary}, related to accelerators, include: 
\begin{itemize}
    \item A multi-TeV muon collider offers a spectacular opportunity in the direct exploration of the energy frontier. Offering a combination of unprecedented energy collisions in a comparatively clean leptonic environment, a high energy muon collider has the unique potential to provide both precision measurements and the highest energy reach in
one machine that cannot be paralleled by any currently available technology. 
\item  Given the LHC results, a 10+ TeV lepton collider going
beyond the classic precision versus energy dichotomy is an ideal machine. 
\item The accelerator challenges of a multi-TeV Muon Collider
are now considered to be overcome based on the technological advances achieved over the past decade. Significant progress has been made in the development of high power targets and of high-field HTS solenoids, in the demonstration of operation of normal conducting RF cavities in magnetic fields, and in the self-consistent lattice designs of the various subsystems. No fundamental show-stoppers
have been identified. Nevertheless, engineering challenges exist in many aspects of the design and targeted R\&D is necessary in order to make further engineering and design progress. 
\item There is an established plan and funding for muon collider related R\&D activities in Europe and it is imperative for Snowmass/P5 to reestablish R\&D efforts in the US and to enable participation of US physicists in the International Muon Collider Collaboration (IMCC). 
\item The most fruitful path forward towards the development of a conceptual design of a Muon Collider would be the engagement of the U.S. community in the IMCC. The U.S. Muon Collider community is well positioned to provide crucial contributions to physics studies, further advance
the accelerator technology and detector instrumentation, and explore options for domestic siting of a muon collider. An Integrated National Collider R\&D Initiative discussed in Snowmass \cite{IFC} can  provide a much needed platform for R\&D funding for such accelerator and detector development.
\item Fermilab could be considered a candidate site for a Muon Collider with a center-of-mass energy reach at the desirable 10-TeV scale. The synergy with the existing/planned accelerator complex and neutrino physics program at FNAL is an additional stimulus for such investment of effort. A set of Muon Collider design options, with potential siting at FNAL, could be a goal contribution for discussions at the IMCC and the international committees to eventually form a global consensus decision on siting and selection of the Muon Collider. Having a pre-CDR document summarizing design for the FNAL-sited Muon Collider in time for the next Snowmass is a good goal. The preparation of such a document will require substantial, yet affordable, investment. Such an investment will reinvigorate the US high-energy collider community and enable much needed global progress towards the next energy frontier.
\end{itemize}

\section{Implementation Task Force (ITF) analysis}
\label{ITF}

The AF Implementation Task Force (ITF) was organized and charged with developing metrics and processes to facilitate a comparison between projects. The ITF is chaired by Thomas Roser (BNL) and comprises 11 additional world-renowned accelerator experts from Asia, Europe and the US, including two members of the Snowmass Young(the Snowmass'21 organization of early career researchers), and three EF and TF liaisons. Corresponding metrics have been developed for uniform comparison of the proposals ranging from Higgs/EW factories to multi-TeV lepton, hadron and $ep$ collider facilities, based on traditional and advanced acceleration technologies. An additional group consisted of versions of the proposals that could be located at FNAL. More than 30 collider concepts have been comparatively evaluated by the ITF in terms of physics reach (impact), beam parameters, size, complexity, power, environment concerns,  technical risk, technical readiness, validation and R\&D required, cost and schedule. The ITF report  documents the metrics and processes, and presents comparative evaluations of future colliders \cite{ITF}. 
 
%

Summary Table \ref{tab:ITF} is a compressed version of the ITF report Tables 1-5 (see \cite{ITF}) and lists the main parameters along with four columns with a summary value for technical risk (years of pre-project R\&D needed), technically limited schedule (years until first physics), project costs (2021 B\$ without contingency and escalation), and environmental impact (the most important impact is the estimated operating electric power consumption). The significant uncertainty in these values was addressed by giving a range where appropriate. 

The years of required pre-project R\&D is just one aspect of the technical risk, but it provides a relevant and comparable measure of the maturity of a proposal and an estimate of how much R\&D time is required before a proposal could be considered for a project start (CD0 in the US system). Pre-project R\&D includes both feasibility R\&D, R\&D to bring critical technologies to a technical readiness level (TRL) of 4-5, as well as necessary R\&D to reduce cost and electric power consumption. The extent of the cost and power consumption reduction R\&D is not well defined and it was assumed that it can be accomplished in parallel with the other pre-project R\&D. Nevertheless this R\&D is likely most important for the realization of any of these proposals. (Note that by using the proponent-provided luminosity values, ITF chose not to evaluate the risk of not achieving this aspect of performance. However,  performance risk was included in the evaluation of technical readiness in the ITF report.) The time to first physics in a technically limited schedule is most useful to compare the scientific relevance of the proposals. It includes the pre-project R\&D, design, construction and commissioning of the facility.

The total project cost follows the US project accounting system {\it but without escalation and contingency.} Various parametric models were used by ITF to estimate this cost, including the cost estimated by the proponents. The cost estimate uses known costs of existing installations and reasonably expected costs for novel equipment. For future technologies, pre-project cost reduction R\&D may further reduce the cost estimates used by the ITF.

Finally the electric power consumption is for a fully operational facility including power consumption of all necessary utilities. The ITF used the information from the proponents if they provided it, otherwise it made  rough estimates based on expert judgement. Pre-project R\&D to improve energy efficiency and to develop more energy efficient accelerator concepts, such as energy recovery technologies, have the potential to reduce the electric power consumption significantly from the values listed in the tables.

Any of the future collider projects will constitute one of the largest - if not the largest - science facilities in particle physics. The cost, the required resources and, maybe most importantly, the environmental impact in the form of large energy consumption will approach or exceed the limit of affordability. The ITF suggests that the Snowmass'21 final report recommends that  R\&D to reduce the cost and the energy consumption of future collider projects is given high priority.

\begin{table}[b!]
\footnotesize
\centering
\begin{tabular}{| l | c | c | c | c | c | c |}
\hline
\hline
 Proposal Name & c.m. energy & Luminosity/IP & Yrs. pre- & Yrs. to 1st & Constr. cost & Electr. power \\ 
  & [TeV] & [10$^{34}$ cm$^{-2}$s$^{-1}$] & project R\&D & physics & [2021 B\$] & [MW] \\  
\hline
FCC-ee$^{1,2}$ & 0.24 & 7.7 (28.9) & 0-2 & 13-18 & 12-18 & 290 \\  
\hline 
CEPC$^{1,2}$ & 0.24 & 8.3 (16.6) & 0-2 & 13-18 & 12-18 & 340 \\  
\hline 
ILC$^{3}$-0.25 & 0.25 & 2.7  & 0-2 & $<$12 & 7-12 & 140 \\  
 \hline 
 CLIC$^{3}$-0.38 & 0.38 & 2.3  & 0-2 & 13-18 & 7-12 & 110 \\  
\hline
CCC$^{3}$  & 0.25 & 1.3  & 3-5 & 13-18 & 7-12 & 150 \\  
\hline
HELEN$^{3}$ & 0.25 & 1.4  & 5-10 & 13-18 & 7-12 & ~110 \\  
 \hline 
FNAL $e^+e^-$ circ. & 0.24 & 1.2  & 3-5 & 13-18 & 7-12 & ~200 \\ 
\hline
CERC$^{3}$ & 0.24 & 78  & 5-10 & 19-24 & 12-30 & 90 \\  
 \hline 
ReLiC$^{1,3}$  & 0.24 & 165 (330)  & 5-10 & $>$25 & 7-18 & 315 \\  
 \hline 
ERLC$^{3}$  & 0.24 & 90  & 5-10 & $>$25 & 12-18 & 250 \\  
 \hline 
XCC $\gamma \gamma$ & 0.125 & 0.1  & 5-10 & 19-24 & 4-7 & 90 \\  
 \hline 
$\mu\mu$-Higgs & 0.13 & 0.01  & $>$10 & 19-24 & 4-7 & 200 \\  
\hline
\hline
ILC-3 & 3 & 6.1 & 5-10 & 19-24 & 18-30 & $\sim$400 \\  
 \hline 
CLIC-3 & 3 & 5.9  & 3-5 & 19-24 & 18-30 & $\sim$550 \\  
\hline 
CCC-3 & 3 & 6.0 & 3-5 & 19-24 & 12-18 & $\sim$700 \\  
 \hline 
ReLiC-3 & 3 & 47(94)  & 5-10 & $>$25 & 30-50 & $\sim$780 \\  
 \hline 
$\mu\mu$Collider$^{1}$-3 & 3 & 2.3(4.6) & $>$10 & 19-24 & 7-12 & $\sim$230 \\  
 \hline 
LWFA-LC-3 & 3 & 10  & $>$10 & $>$25 & 12-80 & $\sim$340 \\  
 \hline 
PWFA-LC-3 & 3 & 10  & $>$10 & 19-24 & 12-30 & $\sim$230 \\  
 \hline 
SWFA-LC-3 & 3 & 10  & 5-10 & $>$25 & 12-30 & $\sim$170 \\  
\hline
 \hline 
FNAL$\mu\mu$$^{1}$ & 6-10 & 20(40)  & $>$10 & 19-24 & 12-18 & $\sim$300 \\ 
 \hline 
LWFA-LC-15   & 15 & 50  & $>$10 & $>$25 & 18-80 & $\sim$1030 \\  
 \hline 
PWFA-LC-15 & 15 & 50  & $>$10 & $>$25 & 18-50 & $\sim$620  \\  
 \hline 
SWFA-LC-15 & 15 & 50  & $>$10 & $>$25 & 18-50 & $\sim$450  \\  
 \hline 
FNAL $pp$ circ. & 24 & 3.5(7)  & $>$10 & $>$25 & 18-30 & $\sim$400 \\  
\hline
FCC-hh$^{1}$ & 100 & 30(60) & $>$10 & $>$25 & 30-50 & $\sim$560 \\  
\hline 
SPPS$^{1}$& 125 & 13(26)  & $>$10 & $>$25 & 30-50 & $\sim$400 \\  
\hline 
\hline 
LHeC & 1.2 & 1 & 0-2 ? & 13-18 & $<$4 & $\sim$140 \\  
\hline 
FCC-eh & 3.5 & 1  & 0-2 ? & $>$25 & $<$4 & $\sim$140  \\  
\hline 
 CEPC-SPPC-ep & 5.5 & 0.37  & 3-5 & $>$25 & $<$4 & $\sim$300 \\  
\hline 
\hline 
\end{tabular}
\caption{Main parameters of the collider proposals evaluated by the ITF: Higgs/EW factories, multi-TeV lepton collider proposals (3 TeV c.m.e. options), colliders with 10 TeV or higher parton c.m.e.,  and   the lepton-hadron collider proposals. The superscripts next to the name of the proposal in the first column indicate (1) total peak luminosity for multiple IPs is given in parenthesis; (2) energy calibration possible to 100 keV accuracy for $M_Z$ and 300 keV for $M_W$; (3) collisions with longitudinally polarized lepton beams have substantially higher effective cross sections for certain processes. The relevant energies for the hadron colliders are the parton c.m. energy, which can be substantially less than hadron c.m. energy quoted in the table. For each proposal, the ITF estimates are given on the years of pre-project R\&D, years to first physics after decision to proceed, construction cost (including explicit labor, no escalation and no contingency), and facility electric power consumption (adapted from \cite{ITF}).} 
\label{tab:ITF}
\end{table}


\section{Integrated Future Collider R\&D Program proposal}
\label{IFCsumm}

It is widely recognized that future colliders are an essential component of a strategic vision for particle physics.  Conceptual studies and technical developments for several exciting future collider options are underway internationally.   The prevailing view of the global HEP community is that the next large collider facility should be an $e^+e^-$ collider as a Higgs/EW factory.  
The physics case for such a collider is compelling, because it would enable the use of ``the Higgs boson as a new tool for discovery,'' as envisioned in the 2014 P5 report.  The community also recognizes that a collider beyond the capabilities of the HL-LHC, with energy reach to explore the $\sim10$~TeV scale, such as a $\sim 100$~TeV hadron collider or a $\geq10$ TeV muon collider, will be necessary. 

\begin{figure}
\begin{center}
\includegraphics[width=0.95\textwidth]{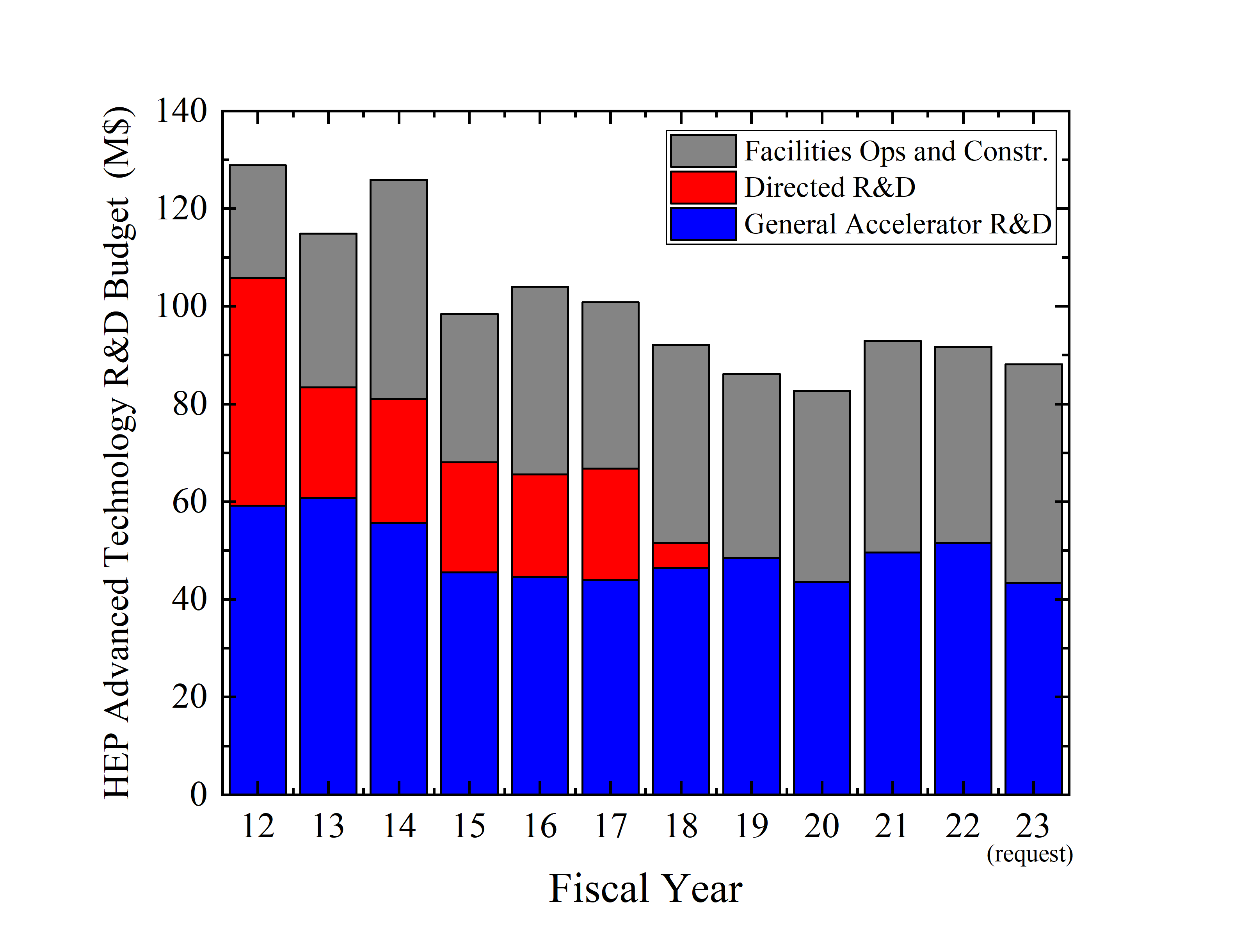}
\caption{Enacted US DOE OHEP budgets for general and directed accelerator R\&D and corresponding test facilities operation and construction (FY12-FY22, in then USD, and FY23 budget request).}
\label{GARDbudgets}
\end{center}
\end{figure}

In order to realize a future collider, a concerted accelerator R\&D program is required \cite{IFC}.   The U.S. HEP accelerator R\&D program currently {\bf has no direct effort in a collider-specific R\&D area.} This gap -- see Fig.\ref{GARDbudgets} -- compromises worldwide progress on promising concepts, compromises the position of the U.S. as a leader in collider design and development, and limits national aspirations for a U.S. leadership role in particle physics. 

In the course of the Snowmass'21 discussions, a new national integrated future colliders R\&D program (IFC) in the DOE Office of High Energy Physics (OHEP) was proposed \cite{IFC}. The overarching objective of the proposed program is to address in an integrated fashion the technical challenges of promising future collider concepts, particularly those aspects of accelerator design, technology, and beam physics that are not covered by the existing General Accelerator R\&D (GARD) program. \textbf{The program will enable; (a) synergistic U.S. engagement in ongoing global efforts (e.g., FCC, ILC, IMCC) and (b) developing collider concepts and proposals for options feasible to be hosted in the U.S.}

The goal of the program is to inform decisions in down-selecting among (now numerous) collider concepts by the next European strategy update in 2025 - 2026, and the next Snowmass and P5 ca. 2030, to help move towards realization of the next collider as soon as possible and subsequently to advance towards a collider at a higher energy scale. 


The characteristics of the proposed IFC program can be summarized as follows:
\begin{itemize}
    \item Scope: sharply focused on future colliders; spans accelerator design, technology and full concept development; complements the existing HEP GARD program; multifaceted but selective and synergistic; integrates all critical R\&D for a concept; priorities guided by P5;
    \item Organization: coherent national program; collaborative effort of U.S. national labs and universities; 
    \item Coordination: centrally coordinated and funded; coordinated with global design studies and R\&D; periodic assessment.
\end{itemize}

The proposed R\&D program would facilitate the realization of future collider facilities, thereby ensuring the continuation of the fruitful endeavors of HEP in advancing the frontiers of our knowledge of the universe. It will also ensure the critical recruitment, development, and retention of a skilled workforce in accelerator science and technology.   


\section{Accelerator technologies and R\&D}

The cost of large accelerators is set by the scale (energy, length, power) and technology. Typically, accelerator components (Normal Conducting (NC) or/and Superconducting (SC) magnets and RF systems) account for $50\pm10 \%$ of the total cost, while the civil construction takes $35\pm15 \%$, and power production, delivery and distribution technology adds the remaining $15\pm10 \%$ \cite{shiltsev2014cost}. While the last two parts are mostly determined by industry, the magnet and RF technologies are linchpins in the progress of accelerators and without R\&D the fraction of technological costs would be much higher.

To develop the necessary technology, as noted in Section 7.1.3, the US DOE OHEP General Accelerator R\&D program supports R\&D in five main thrusts: accelerator and beam physics (ABP), RF acceleration technology (NC and SC RF), magnets and materials, advanced acceleration concepts (AAC), and particle sources and targets. These programs and corresponding facilities operations were supported with a total of roughly 90M\$ a year (see Fig. \ref{GARDbudgets}), distributed as estimated from FY19 as 36\% for AAC, 22\% for RF, 19\% for materials and magnets, 15\% for ABP, and 2\% for targets and sources.

\subsection{Superconducting magnet R\&D}

Superconducting magnets are required for many of the proposed facilities, in some cases with operational parameters well beyond current state-of-the-art, such as muon colliders or next generation high energy
hadron colliders. This need was recognized by the last P5 process and the US Department of Energy-Office of High Energy Physics initiated the US Magnet Development Program (MDP) \cite{gourlay2016us}; \cite{ambrosio2022strategic}. It is a general R\&D
program that pulls together US HEP magnet research groups at DOE laboratories and Universities under a common collaboration, with a focused mission and goals. MDP is closely collaborating with advanced magnet programs worldwide and has developed growing synergies with the NSF-funded National High Magnetic Field Laboratory; with other DOE offices, in particular, DOE-OFES; and with industry, both for conductor development and through the SBIR program. In the EU, the Particle Physics Strategy update process resulted in several recommendations for investments in critical technologies for future colliders and has resulted in the launching of the European High Field Magnet program (HFM) \cite{vedrine2022high}. In Japan, superconducting accelerator magnet programs have been conducted at KEK for more than 40 years and have contributed to several international projects. In China, the primary focus is on Fe-based 12T magnets and high field hybrids utilizing Nb$_3$Sn and HTS.
Superconducting Nb-Ti magnets used in the LHC operate close to the conductor limit at 8.3T. Dipole fields more than 16 T (up to 24 T) are needed, and solenoids, primarily for the muon collider, up to 50 T \cite{alexahin2022critical}. A recently tested US MDP prototype Nb$_3$Sn magnet has reached 14.5 T \cite{zlobin2021ieee} - see Fig.7-6. The fastest rapid cycling magnets (required for a muon collider) have demonstrated up to 300 T/s ramp rates in an HTS-based superconducting magnet test at Fermilab \cite{piekarz2022fast}.
Fields above 16 T require the use of high temperature superconductors (HTS); Bi-2212, REBCO and Fe-based, either in lieu of or combined in a hybrid scheme with Nb$_3$Sn. 

Cost is a particular issue with all viable conductors beyond Nb-Ti. High cost slows R\&D progress and would have a large impact on the cost of a future collider. Conductor cost reduction might be achieved through a combination of continued R\&D specific to cost reduction and development of industrial infrastructure through the recently launched Accelerator R\&D and Production (ARDAP) Program.  Also, there are already co-funded projects with Fusion Energy Sciences (FES, e.g. the Test Facility Dipole) and FES and fusion companies may be significant future magnet drivers.
Superconducting detector magnets are a key component of particle physics experiments and have made significant progress over the past few decades. Further improvements will be required to cope with larger size, radiation loads and physics performance requirements.

A sizeable international magnet R\&D effort over many years will be necessary to achieve the performance levels desired for many of the collider proposals now on the table. Historically, an international collaborative framework has proven to be critical in this area of technology.

As the HEP community explores its best options to enable future discoveries, magnet technology considerations are essential for informed decisions on the feasibility and cost of achieving the physics goals.

Successful projects have benefited from a combination of fundamental R\&D by GARD and directed R\&D. One example is the LHC Accelerator Research Program (LARP) that was started in 2004 as a directed R\&D effort to develop Nb$_3$Sn IR Quadrupoles for the High Luminosity LHC. By 2013, it had achieved its main goals leading to the US LHC Accelerator Upgrade Project.

To maintain this highly successful approach and meet future challenges a new program has been proposed.  The Leading-Edge technology And Feasibility-directed (LEAF) Program \cite{ambrosio2022white} would have a goal to achieve readiness for a future collider decision in the next decade.

\begin{figure}[htbp]
\centering
\includegraphics[width=0.75\linewidth]{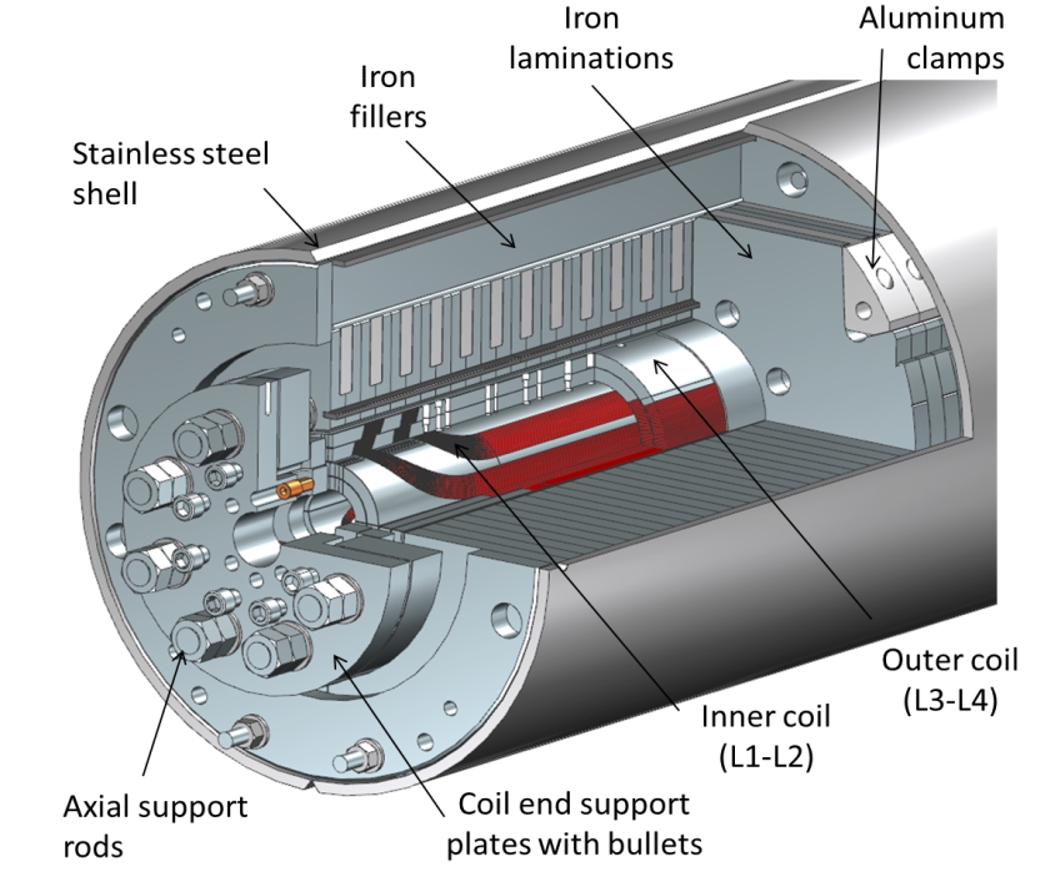}
\caption{15 T Nb$\_3$Sn Magnet.} 
\label{Fig5}
\end{figure}

\subsubsection{AF7 ("Accelerator Technology - Magnets"): main findings and recommendations}
\label{AF7MAGsumm}

The U.S. Magnet Development Program (US-MDP) focuses on fundamental accelerator magnet R\&D providing key magnet technologies which benefit all future accelerators.

\begin{itemize}

\item A successful ongoing national program with short- and long-term goals and plans, efficient management structure, established staff, and facilities.

\item Well-recognized internationally, actively participate in international magnet R\&D activities. 

\item Superconducting magnet technology is critical to most of the proposed colliders and new experiments and strong engagement from the magnet community on future collider roadmaps is critical to ensure success.
\end{itemize}

Discussions and initiatives emerging from the Snowmass’21 process motivate enhanced scope and significant increase in resources to prepare accelerator magnet technology, e.g. \cite{AF7MAGsummary}:

\begin{itemize}

\item Enhance MDP scope and resources to address general magnet R\&D needs for HEP, including magnet design and short model development for specific accelerator applications, e.g. muon collider

\item In addition, invest in critical technology scale up and cost reduction efforts, and preparation for industrialization, with the goal of providing magnet technology readiness for future projects and a sound basis for future facility specifications and costing.

\item Explore opportunities to reduce power requirements for facility operation

\item Provide support for test facility upgrades and operation

\end{itemize}

\subsection{RF technology R\&D}

Accelerator radio frequency (RF) technology has been and remains critical for modern high energy physics (HEP) experiments based on particle accelerators. Tremendous progress in advancing this technology has been achieved over the past decade in several areas. These achievements and new results expected from continued R\&D efforts could pave the way for upgrades of existing facilities, improvements to accelerators already under construction, well-developed proposals and/or enable concepts under development.

The ongoing RF technology R\&D efforts closely follow the decadal roadmap that was developed in the framework of the DOE General Accelerator R\&D (GARD) program in 2017. The roadmap
reflects the most promising research directions that can potentially enable future experimental HEP programs. Similar to the DOE GARD roadmap, the European particle physics community developed
a roadmap for European accelerator R\&D presented in a report that includes a roadmap for high-gradient RF structures and systems. The two roadmaps cover similar RF technology topics thus presenting opportunities for collaboration between the U.S. and European institutions.

\begin{figure}[htbp]
\centering
\includegraphics[width=0.75\linewidth]{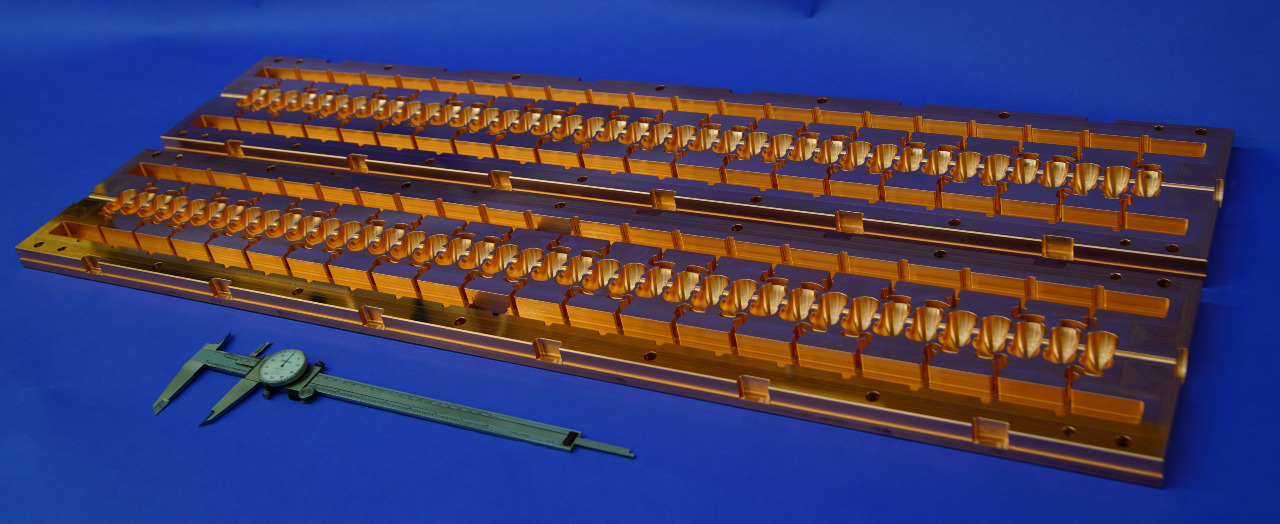}
\caption{Both halves of the C$^3$ prototype structure prior to braze. The one meter structure consists
of 40 cavities. A RF manifold that runs parallel to the structure feeds 20 cavities on each side. The structure operates at 5.712 GHz - from \cite{nanni2022c}} 
\label{Fig6}
\end{figure}

The highest gradient large-scale NC RF system is the 28 MeV/m linac of the SwissFEL at the PSI (Switzerland). Up to 100 MV/m accelerating beam gradients were achieved in the CLIC 12 GHz structures at the CERN test facility, while up to 150 MV/m gradients were reported in the first test of short 11.4 GHz NC structures cooled to 77 K at SLAC \cite{bai2021c} - Fig.\ref{Fig6}. 
As for SRF, the largest scale accelerator to date is the 17.5 GeV 1.3 GHz pulsed linac of the European XFEL at DESY that has achieved an average beam gradient of roughly 23 MV/m with a nominal Superconducting RF (SRF) cavity quality factor of $Q_0\simeq1.4\cdot 10^{10}$ at 2 K. The full ILC specification on the beam acceleration gradient of 31.5 MV/m has been demonstrated at the FNAL FAST facility \cite{broemmelsiek2018record}. Recent advances in SRF cavity processing such as {\it nitrogen doping} allow further improvement of the quality factor to $(3-6)\cdot 10^{10}$ (hence, reducing the required cryogenic cooling power) and an initial cryomodule for the LCLS-II-HE, operated at 25 MV/m with a $Q_0 = 3\cdot 10^{10}$.  Further R\&D aims for 50 MV/m gradients in 1.3 GHz structures \cite{grassellino2017unprecedented}. 
An active ongoing accelerator R\&D program for future multi-MW proton beams includes development of more efficient power supplies with capacitive energy storage and recovery, more economical RF power sources such as 80\% efficient klystrons, magnetrons, and solid-state RF sources  (compared to current $\sim$ 55\% ) see for example \cite{nanni2022c}; \cite{ives2022high}.

\subsubsection{AF7 ("Accelerator Technology - RF"): main findings and recommendations}
\label{AF7RFsumm}

The GARD roadmap continues to serve as community-developed guidance for RF technology R\&D, but based on recent progress it would benefit from some mid-course corrections. Based on the discussions and submitted White Papers, the following are key directions that should be pursued during the next decade \cite{AF7RFsummary}: 

\begin {itemize}

\item Studies to push performance of niobium and improve our understanding of SRF losses and ultimate quench fields via experimental and theoretical investigations. 
\item Developing methods for nano-engineering the niobium surface layer and tailoring SRF cavity performance (shunt impedance, Q-factor and gradient) to a specific application, e.g., a linear collider, a circular collider, or a high intensity proton linac.
\item Investigations of new SRF materials beyond niobium via advanced deposition techniques and bringing these materials to the point where they can be used in accelerators.
\item Developing advanced SRF cavity geometries to push accelerating gradients of bulk niobium cavities to approximately 70 MV/m for either an upgrade of the ILC or compact SRF linear collider.
\item Pursuing R\&D on companion RF technologies to mitigate field emission, provide precise resonance control, etc. 
\item Research on application of SRF technology to dark sector searches, such as axions. 
\item R\&D on high-gradient normal conducting RF structures operating at cryogenic temperatures with a gradient of greater than 150 MV/m as a promising way toward a compact linear collider. 
\item R\&D on high frequency and multi-frequency structures to transcend limits on shunt impedance and accelerating gradients.
\item Investigation of novel materials and manufacturing techniques to improve high gradient performance and remove design constraints.
\item Developing high efficiency, low-cost RF sources that would benefit many operating and practically every future intensity or energy frontier machine.
\item Studies dedicated to industrialization and cost reduction of fabricating RF components and systems.
\item Continue research on advanced SWFA structures to bring them closer to practical applications. 
\item Experimental and theoretical research to further our understanding of the RF breakdown physics.
\item Continued development of computational tools for multi-physics and virtual prototyping.
\item Developing a community ecosystem for accelerator and beam physics modeling that would incorporate a comprehensive set of high-performance simulation tools for RF-based accelerators.

\end {itemize}

To support these key research directions, there is a need to upgrade the existing or build new R\&D and test facilities that have the capabilities to investigate the new concepts and help integrate them into systems with ready access to researchers. 

\subsection{Targets and sources R\&D}
The next generation of high power targets for future accelerators will use more complex geometries, novel materials, and new concepts (like flowing granular materials \cite{ammigan2022novel}). Under active development are advanced numerical approaches that satisfy the physical design requirements of reliable beam targets \cite{barbier2022modeling}. In parallel, there is work to be done on radiation hardened beam instrumentation \cite{yonehara2022radiation}, development of irradiation methods for high power targets and irradiation facilities \cite{pellemoine2022irradiation}.

There are challenges in high-intensity high-brightness beam sources for future accelerators. They are particularly formidable for sources of beams with special characteristics, such as polarized electrons and ions or ultra-small emittances, and for  tertiary and secondary particles, such as muons and positrons. In the longer run, high intensity positron sources, that are critical for  future $e^+e^-$ linear colliders, may be simplified using advanced
accelerator concepts and photon-driven schemes that can potentially outperform conventional $e^+$ production techniques \cite{chaikovska2022positron}.

\subsubsection{AF7 ("Accelerator Technology - Targets and Sources"): main findings and recommendations}
\label{AF7SRCTARGsumm}

Key recommendations of the Snowmass'21 AF7 topical group ("Accelerator Technology - Targets and Sources") include \cite{AF7TSsummary}: 
\begin{itemize}
\item Develop or upgrade Post-Irradiation Examination (PIE) facilities that will allow testing activated novel materials.
\item Develop and validate alternative beams that mimic the effects of high-energy proton irradiation.
\item Develop irradiation stations relevant to assess radiation damage in various materials.
\item Develop and validate numerical tools to predict radiation damage.
\item Develop novel concepts and novel target materials.
\item Develop radiation-resistant sensors for better numerical validation of the structural and thermal simulations.
\item Strengthen the high-power target community through the RaDIATE collaboration and High-Power Targetry workshops.
\item Leverage the synergy with other communities facing similar challenges (light source facilities, Accelerator Driven Systems and fusion).
\item Advance the state of development of a 4th generation Electron Cyclotron Resonance Ion Source (ECRIS) that can be achieved using higher magnetic fields.
\item Accelerate development of high intensity positron sources including the use of advanced accelerator concepts.
\item Continue development of high brightness, high average current sources (DC, RF and SRF guns) and explore applying machine learning to advance source performance and stability.
\item R\&D in high current electron guns is needed for efficient charge breeding of high-intensity beams.
\item Development of cathode testing facilities is critical.

\end{itemize}


\subsection{Advanced acceleration concepts}


{ Wakefields driven in plasmas or advanced structures offer high accelerating gradients, and R\&D is in progress to develop future collider options.}
Plasma waves can be effectively excited by short and powerful external drivers. A benefit of plasma acceleration is that plasma is not limited by the conventional RF breakdown processes as, in some sense, it is already broken down and thus acceleration gradients of many GV/m are possible.  Three effective ways to excite plasma have been demonstrated in the past two decades: i) by intense electron bunches -- 9 GeV acceleration over 1.3m at $10^{17}$cm$^{-3}$ has been achieved at the SLAC FACET facility \cite{litos20169}; ii) by short laser pulses -- an electron beam energy gain of up to 7.8 GeV over 20 cm in a few $10^{17}$cm$^{-3}$ plasma observed at the LBNL BELLA facility \cite{gonsalves2019petawatt}; and iii) by self-modulated high energy proton bunches -- 2 GeV gain over 10m at $10^{15}$cm$^{-3}$  reported by the CERN AWAKE experiment \cite{adli2018acceleration}.  

There are three remarkable recent developments on the way to practical plasma-based beam accelerators: i) EuPRAXIA has been included in the ESRFI 10-20 yrs roadmap \cite{assmann2020eupraxia} (EuPRAXIA is a 569 MEuro European plasma accelerator project {\it proposal}, supported by 50 institutions from 15 countries, and aiming for 5 GeV electron beam acceleration and development of plasma-based FELs); ii) the laser wakefield accelerator at SIOM/CAS in Shanghai (China) has achieved an 
accelerated 0.5 GeV electron beam (produced by a 200 TW laser exciting plasma in a 6 mm He gas jet) sufficient for initiation of the FEL generation of 27 nm light in the downstream undulators --  making it the first demonstration of a plasma-based FEL \cite{wang2021fel}; and iii) a similar free electron lasing demonstration at the LNF-INFN with a beam-driven plasma accelerator \cite{shpakov2021first}. 


In addition to plasma acceleration, relativistic electrons or positrons passing near a material boundary will produce wakefields when the particle velocity exceeds the Cherenkov radiation condition. The longitudinal wakes can be used to accelerate an appropriately phased trailing beam \cite{gai1988experimental} or can be extracted by a high efficiency RF coupler as a high power RF source \cite{gao2008design}.  The CLIC linear collider design is an example of using 'structure wakes' to power a linac.  There has also been significant progress on another approach that utilizes laser-driven microstructure accelerators \cite{england2022laser}.

Structure wakefield acceleration (SWFA) has an advantage for application to linear colliders because the structures naturally accelerate electrons and positrons and are expected to have smaller challenges preserving the beam emittance than plasma accelerators. Recent progress in development has increased the gradient to 300 MV/m and wakefield power generation to 500 MW \cite{jing2022continuous}. The next proposed steps are to design, construct and test a laboratory-scale SWFA module and incorporate it into a SWFA-based facility design with the intent of implementing a dedicated test facility as a scalable demonstration of the technology aimed at an energy frontier machine.

In principle, wakefield-based linear accelerators can be employed in high energy $e^+e^-$/$\gamma \gamma$ colliders, but a significant R\&D effort is required to address many critical issues, such as acceleration of positrons, staging of multiple plasma acceleration cells, power efficiency, emittance control, jitter and scattering in plasma, beamstrahlung, etc \cite{shiltsev2021modern}, \cite{CERN2022RnD}.  New concepts continue to emerge that may address known challenges or further expand capabilities.


\subsubsection{AF6 (''Advanced Acceleration Concepts"): main findings and recommendations}
\label{AF6summ}

The AF6 Topical Group focused on understanding the limits, the state of the design concepts, and the status of the R\&D for plasma and structure wakefield acceleration.  At this time, there are no parameter sets for a plasma or structure based linear collider that self-consistently address the known accelerator physics challenges -- see Appendix 7.6 of Ref. \cite{ITF} for the latest suggested parameters.  However, acceleration of beams with ultra-high gradients (GV/m and beyond) will reduce the dimensions and thus potentially reduce the costs and power consumption of future high energy physics machines with the potential to offer $e^+e^-$ and $\gamma\gamma$ colliders at and beyond 15 TeV. Recognizing this promise, the last Snowmass and P5/HEPAP recommended, and DOE developed with the community, an organized Advanced Accelerator Development Strategy, and work has been aligned to this strategy.

Over the last 15 years, three major facilities for advanced accelerator research have been constructed and operated in the US: FACET, BELLA, and AWA.  A number of important milestones have been achieved including multi-GeV acceleration in a single stage, limited positron acceleration, efficient local loading of the structure, the first staging of plasma accelerators, and demonstration of beam shaping to improve efficiency in plasmas and structures. Details, including references, describing the status and needs of the field can be found in the AF6 Topical Group Summary \cite{AF6summary}.

In order to move forward, a vigorous R\&D program will be required. The US is in a good position in this respect with several state-of-the-art of beam test facilities mainly dedicated to research in the advanced accelerator field, including FACET-II, BELLA, ATF, AWA and FAST-IOTA as well as numerous universities. As described, vigorous R\&D using these facilities and programs is needed to push forward the key next steps in the Development Strategy including staging of multiple modules at multi-GeV, high efficiency stages, preservation of emittance for electrons and positrons and shaped beams, as well as the development of efficient, low-cost and high repetition rate drivers. 

The development of self-consistent parameter sets for compact high-gradient colliders addressing the known physics challenges is critical to guide the efforts and provide a clear and actionable R\&D path for an HEP facility. Studies should provide detailed examples of how the main challenges can be addressed and clarify where experimental demonstrations, or detailed simulations of the relevant sub-systems, are needed. In addition, cost estimates, including the impact of some of the stringent component tolerances, should be made to understand if the potential cost benefit of the new technologies can be realized. 

Developing the parameters and pursuing the R\&D may require funding that has not been available to date. The European roadmap for accelerators \cite{CERN2022RnD} includes a full chapter on advanced accelerators and is well-aligned with our call for an organized integrated design study. Even though the advanced accelerator field was born and is still squarely centered in the US, it is telling that the most recent high profile plasma-based FEL demonstrations occurred in Europe and Asia. This kind of research in the US is unfortunately not seen as directly impacting HEP, putting in jeopardy the US leadership in the field. 

The explicit suggestions from the AF6 working group \cite{AF6summary} focus on priority research to continue to address and update the Advanced Accelerator Development Strategy:
\begin{itemize}

\item Vigorous research on advanced accelerators including experimental, theoretical, and computational components should be conducted as part of the General Accelerator R\&D program. This will advance the advanced accelerator R{\&}D roadmaps towards future high energy colliders, develop intermediate applications, and ensure international competitiveness. Priority directions include staging of multiple modules at multi-GeV, high efficiency stages, preservation of emittance for electrons and positrons, high fidelity phase space control, active feedback precision control, and shaped beams and deployment of advanced accelerator applications. 

\item A targeted R\&D program addressing high energy advanced accelerator-based colliders (e.g. to 15~TeV, with intermediate options) should develop integrated parameter sets in coordination with international efforts. This should detail components of the system and their interactions, such as the injector, drivers, plasma source, beam cooling, and beam delivery system. This would set the stage for an integrated design study and a future conceptual design report, after the next Snowmass.  

{ \item Research in near-term applications should be recognized as essential to, and providing leverage for, progress towards HEP colliders. The interplay and mutual interests in this area between Offices in DOE-SC including HEP, BES, FES and ARDAP as well as with NSF, NNSA, defense and other agencies should be strengthened to advance and leverage research activities aimed at real-world deployment of advanced accelerators.}

{ \item Advanced accelerators should continue to play a key-role in workforce development and diversity in accelerator physics. University programs and graduate students greatly benefit from the scientific visibility of the advanced accelerator field. Access to user facilities for graduate students and early career researchers as well as formal and hands on training opportunities in advanced beam and accelerator physics should be continued and enhanced.}

\item Enhanced driver {R\&}D is needed to develop the efficient, high repetition rate, high average power laser and charged particle beam technology that will power advanced accelerator colliders and societal applications. 
 
\item Support of upgrades for Beam Test Facilities are needed to maintain progress on advanced accelerator Roadmaps. These include development of a high repetition rate facility, proposed as kBELLA, to support precision active feedback and high rate; independently controllable positrons to explore high quality acceleration, proposed at FACET-II; and implementation of a integrated SWFA demonstrator, proposed at AWA.

\item A study for a collider demonstration facility and physics experiments at an intermediate energy (c.a. 20--80~GeV) should establish a plan that would demonstrate essential technology and provide a facility for physics experiments at intermediate energy.  

\item A DOE-HEP sponsored workshop in the near term should update and formalize the U.S. advanced accelerator strategy and roadmaps including updates to the 2016 AARDS Roadmaps, and to coordinate efforts.


\end{itemize}
\subsection{Accelerator and Beam Physics and Accelerator Education}

Accelerator and beam physics (ABP), the science of charged particle beams, is an essential aspect of designing and building the next frontier accelerators. A robust and scientifically challenging program in accelerator and beam physics is critical for the field of particle physics to be productive and successful. Major ABP challenges aim at improving the reliability, performance, safety, and cost reduction of future accelerators and push the envelope of beam intensity, beam quality, beam measurement and control, and development of methods to model and predict beam behavior.

There are many notable recent developments in the physics of beams. For example, several novel cooling schemes were experimentally demonstrated at operational accelerators. A true breakthrough was the demonstration of the {\it ionization cooling} of 140 MeV/c muons at the MICE experiment at RAL (UK) -- some 
10\% beam emittance reduction was observed in a single pass through the cooling section \cite{mice2020demonstration}. In 2020, {\it “bunched” electron beam cooling} of ions in RHIC ($\gamma \sim 5$) -- remarkable by the pioneering use of high quality bunched electron beams from an electron beam RF photoinjector gun (before, only DC electron accelerators were used with limited capability to get to very high energies) -- was demonstrated at BNL \cite{fedotov2020experimental}. In 2021 the Fermilab team successfully carried out a proof-of-principle experiment on the {\it optical stochastic cooling} of 100 MeV electrons in the IOTA ring in which the use of undulator magnets - instead of electrostatic pickups in traditional stochastic cooling setups - expanded the feedback system bandwidth by several orders of magnitude to the THz range \cite{jarvis2022first}. 

Efforts have been ramping-up in recent years to use Artificial Intelligence and Machine Learning (AI/ML) to enhance the performance of accelerators and beamlines \cite{duris2020bayesian, arpaia2021machine}. There is significant promise of applications of AI/ML in beam diagnostics, controls, and modeling \cite{edelen2016neural}. Opportunity exists in broadening AI/ML methods for early detection of a broad range of accelerator component or subsystem failures \cite{edelen2016first} and for optimization of advanced numerical simulations through identification of the most promising combinations of parameters thereby reducing the total number of required simulations \cite{koser2021input}.

Future circular and linear $e^+e^-$ colliders require collision optimization studies including the 3D beam size flip-flop from the beam-beam effect, large dynamic aperture Interaction Region (IR) design; pico-meter vertical emittance preservation techniques in high-charge circular colliders with strong focusing IRs, effects of detector solenoids, and beam-beam effects; end-to-end emittance preservation simulations for linear colliders augmented with experimental tests of beam-based alignment techniques in the presence of realistic external noise sources; plasma-lens-based final focus and beam transport system designs \cite{ariniello2019transverse, lindstrom2021staging}. 

Required ABP explorations for future hadron colliders include 
efficient collimation techniques \cite{scandale2019channeling}, electron lenses for Landau damping and collimation \cite{shiltsev2021electronlens}, dynamic aperture optimization methods to make possible new integrable optics solutions \cite{danilov2010nonlinear}, and studies to obtain lower emittances from new particle sources for injecting beams in high-bunch-charge colliders.


A recent community exercise summed up the needs and directions of future developments in ABP \cite{nagaitsev2021accelerator}. Among many, those include issues related to beam loss control in high-intensity high-power accelerators (space-charge effects, instabilities, collimation, electron lens compensation, integrable optics, etc) which require innovative approaches, theoretical and experimental studies (at, e.g., the IOTA ring \cite{antipov2017iota}, and operational accelerators in the US and abroad) and validated computer models/codes. A key challenge would be to reduce particle losses $(dN/N)$ at a faster rate than increases in achieved beam intensity (power) $(N)$ \cite{shiltsev2020superbeams}. 

Traditionally, HEP has played a leading role in education and workforce training for particle accelerators. Currently, this task is surely shared by other SC offices, e.g., NP (that will need a significant workforce for the EIC) and BES (e.g. for operational and future X-ray sources and spallation neutron sources), that plan building new or upgrading existing facilities in the next decades - all of these need trained and talented accelerator personnel. Naturally, the EIC and other SC accelerator projects will also provide an excellent training ground for personnel that would, in the future, be needed for HEP machines, such as high intensity proton sources and high energy colliders. All that requires a shared vision on accelerator workforce development and R\&D with NP and BES. The DOE OHEP should take a lead in organizing corresponding discussions within the accelerator community and formulation of an integrated vision across the Office of Science and NSF.

\subsubsection{AF1 ("Accelerator and Beam Physics and Accelerator Education"): main findings and recommendations}
\label{AF1summ}
The physics limits of ultimate beams indicate that it will be very difficult to reach an order of magnitude higher energy beyond currently proposed future colliders such as ILC, FCC and CEPC with today’s conventional accelerator technology. Hence, intensified R\&D in advanced beam physics and accelerator technologies is needed to address known challenges and reach ultimate beams. 

On the contrary, support for fundamental beam physics research has been declining. In the 2010's, the NSF reduced its support of Accelerator Science while funding by DOE through GARD and Accelerator Stewardship has been steady in then-year dollars, thereby declining with inflation - see Fig.\ref{GARDbudgets}. With the energy frontier shifted from the US to Europe, collider expertise, particularly in $e^+e^-$ and muon colliders, has not been able to maintain a healthy profile. In addition to slowing advancement, it makes it difficult to maintain a viable R\&D portfolio and threatens student training and work-force development in US accelerator science. Integrated efforts are needed to mitigate this situation and maintain at adequate levels the Beam Physics and Accelerator Science \& Engineering (AS\&E) education and outreach programs in the US. 

A robust and scientifically challenging R\&D program in accelerator and beam physics is needed to position the field of US High Energy Physics to be productive and competitive for decades to come. Corresponding recommendations of the AF1 topical group include \cite{AF1summary}:

\begin{itemize}
\item Continuous reduction of the inflation-adjusted budget of the OHEP GARD program in general and ABP program should be stopped, and the level of support adequate to the challenges ahead of the ABP thrust should be secured; 
\item Establishment of a decadal road map of accelerator and beam physics research in the DOE OHEP General Accelerator R\&D (GARD) program to address the four ABP {\it Grand Challenges} in beam intensity, quality, control and prediction \cite{ABPGC} ;
\item Re-establish a program of beam physics research on general collider related topics towards future $e^+e^-$ and muon colliders;
\item Strengthen and expand capabilities of the US accelerator beam test facilities to maintain their competitiveness with respect to worldwide capabilities. 
\end{itemize}

To build and maintain a strong diverse and inclusive workforce to support future HEP accelerator facilities, we also propose that the community pursue the following efforts to further strengthen Beam physics and Accelerator Science \& Engineering (AS\&E) education and outreach program:
\begin{itemize}
    \item Gather integrated statistics on workforce composition and needs as well as gender, and ethnicity for AS\&E students and workers in the labs, universities, and industry to monitor progress and better guide long-term efforts. This can be achieved by extending roles of the USPAS.
    \item The AS\&E field should organize a yearly undergraduate level recruiting program structured to draw in talent broadly and also enhance recruiting of women and underrepresented minority (URM) students. This could be coordinated with the USPAS.
    \item Increase US Particle Accelerator School (USPAS) office effort by one FTE to extend roles listed above to benefit the community; improve technical IT support of classes; and for long-range planning and stability.  
    \item Strengthen accelerator research at universities by funding professors and projects on campus and in collaborative lab efforts to increase visibility among undergraduates to recruit talent into the field.
    \item Lower the barriers to participation women and URM talent into the field and take concrete measures to improve discourse and support quality of life and family support issues at the labs to broadly retain talent. Recommended steps are detailed in the Education, Outreach, and Diversity section.   
\end{itemize}





\bibliographystyle{JHEP}
\bibliography{Accelerator}

\providecommand{\href}[2]{#2}\begingroup\raggedright\begin{thebibliography}{10}

\bibitem{shiltsev2020particle}
V.~Shiltsev, \emph{Particle beams behind physics discoveries}, {\emph{Physics
  Today} {\bfseries 73} (2020) }.

\bibitem{2014P5}
S.~Ritz, H.~Aihara, M.~Breidenbach, B.~Cousins, A.~de~Gouvea, M.~Demarteau
  et~al., \emph{Building for discovery: strategic plan for {US} particle
  physics in the global context}, {\emph{HEPAP Subcommittee} (2014) }.

\bibitem{broemmelsiek2018record}
D.~Broemmelsiek, B.~Chase, D.~Edstrom, E.~Harms, J.~Leibfritz, S.~Nagaitsev
  et~al., \emph{Record high-gradient {SRF beam acceleration at Fermilab}},
  {\emph{New Journal of Physics} {\bfseries 20} (2018) 113018}.

\bibitem{mice2020demonstration}
M.~Bogomilov et~al., \emph{Demonstration of cooling by the muon ionization
  cooling experiment}, {\emph{Nature} {\bfseries 578} (2020) 53}.

\bibitem{fcchh}
{{The FCC Collaboration}}, \emph{{FCC-hh: The Hadron Collider}},
  {\emph{{Eur.~Phys.~J.~Spec.~Top.}} {\bfseries 228} (2019) }.

\bibitem{zlobin2021ieee}
A.~Zlobin, D.~Turrioni, S.~Krave, S.~Stoynev, S.~Caspi, J.~Carmichael et~al.,
  \emph{Reassembly and test of high-field nb$_3$ sn dipole demonstrator
  {MDPCT1}}, {\emph{IEEE Trans. Appl. Supercond.} {\bfseries 31} (2021)
  4000506}.

\bibitem{GARD2015subpanel}
A.~Discovery, \emph{{Strategic Plan for Accelerator R\&D in the US}},
  {\emph{HEPAP Accelerator R\&D Subpanel Report, April} (2015) }.

\bibitem{shiltsev2020superbeams}
V.~Shiltsev, \emph{Superbeams and neutrino factories—two paths to intense
  accelerator-based neutrino beams}, {\emph{Modern Physics Letters A}
  {\bfseries 35} (2020) 2030005}.

\bibitem{AF2summary}
R.~Zwaska et~al., \emph{{Report of the Snowmass'21 AF2}}, {\emph{arXiv preprint
  arXiv:2209.xxxxx} (2022) }.

\bibitem{jaeckel2020quest}
J.~Jaeckel, M.~Lamont and C.~Vall{\'e}e, \emph{The quest for new physics with
  the {Physics Beyond Colliders} programme}, {\emph{Nature Physics} {\bfseries
  16} (2020) 393}.

\bibitem{beacham2019physics}
J.~Beacham, C.~Burrage, D.~Curtin, A.~De~Roeck, J.~Evans, J.L.~Feng et~al.,
  \emph{Physics beyond colliders at {CERN}: beyond the standard model working
  group report}, {\emph{Journal of Physics G: Nuclear and Particle Physics}
  {\bfseries 47} (2019) 010501}.

\bibitem{budker2022expanding}
D.~Budker, J.C.~Berengut, V.V.~Flambaum, M.~Gorchtein, J.~Jin, F.~Karbstein
  et~al., \emph{Expanding nuclear physics horizons with the {Gamma Factory}},
  {\emph{Annalen der Physik} {\bfseries 534} (2022) 2100284}.

\bibitem{ahdida2019sensitivity}
C.~Ahdida, R.~Albanese, A.~Alexandrov, A.~Anokhina, S.~Aoki, G.~Arduini et~al.,
  \emph{Sensitivity of the {SHiP} experiment to heavy neutral leptons},
  {\emph{Journal of High Energy Physics} {\bfseries 2019} (2019) 1}.

\bibitem{ahdida2019sps}
C.~Ahdida, R.~Alia, G.~Arduini, A.~Arnalich, P.~Avigni, F.~Bardou et~al.,
  \emph{{SPS} beam dump facility--comprehensive design study}, {\emph{arXiv
  preprint arXiv:1912.06356} (2019) }.

\bibitem{pellico2022fnal}
W.~Pellico, C.~Bhat, J.~Eldred, C.~Johnstone, J.~Johnstone, K.~Seiya et~al.,
  \emph{{FNAL PIP-II} accumulator ring}, {\emph{arXiv preprint
  arXiv:2203.07339} (2022) }.

\bibitem{toups2022sbn}
M.~Toups, R.~Van~de Water, B.~Batell, S.~Brice, P.~deNiverville, J.~Eldred
  et~al., \emph{{SBN-BD: O(10 GeV)} proton beam dump at {Fermilab's PIP-II
  Linac}}, {\emph{arXiv preprint arXiv:2203.08102} (2022) }.

\bibitem{apyan2022darkquest}
A.~Apyan, B.~Batell, A.~Berlin, N.~Blinov, C.~Chaharom, S.~Cuadra et~al.,
  \emph{{DarkQuest: A dark sector upgrade to SpinQuest at the 120 GeV Fermilab
  Main Injector}}, {\emph{arXiv preprint arXiv:2203.08322} (2022) }.

\bibitem{akesson2022current}
T.~Akesson, N.~Blinov, L.~Brand-Baugher, C.~Bravo, L.K.~Bryngemark, P.~Butti
  et~al., \emph{Current status and future prospects for the {Light Dark Matter
  eXperiment}}, {\emph{arXiv preprint arXiv:2203.08192} (2022) }.

\bibitem{AF5summary}
E.~Prebys, M.~Lamont and R.~Milner, \emph{{Accelerators for Rare Processes and
  Physics Beyond Colliders: Report of the AF5 Topical Group to Snowmass 2021}},
  {\emph{arXiv preprint arXiv:2209.06289} (2022) }.

\bibitem{shiltsev2021modern}
V.~Shiltsev and F.~Zimmermann, \emph{Modern and future colliders},
  {\emph{Reviews of Modern Physics} {\bfseries 93} (2021) 015006}.

\bibitem{ITF}
T.~Roser et~al., \emph{{Report of the Collider Implementation Task Force}},
  {\emph{arXiv preprint arXiv:2208.06030} (2022) }.

\bibitem{lou2019circular}
X.~Lou, \emph{The circular electron positron collider}, {\emph{Nature Reviews
  Physics} {\bfseries 1} (2019) 232}.

\bibitem{stapnes2019compact}
S.~Stapnes, \emph{The compact linear collider}, {\emph{Nature Reviews Physics}
  {\bfseries 1} (2019) 235}.

\bibitem{michizono2019international}
S.~Michizono, \emph{{The international Linear Collider}}, {\emph{Nature Reviews
  Physics} {\bfseries 1} (2019) 244}.

\bibitem{bai2021c}
M.~Bai, T.~Barklow, R.~Bartoldus, M.~Breidenbach, P.~Grenier, Z.~Huang et~al.,
  \emph{C$^{3}$: A "cool" route to the {Higgs} boson and beyond}, {\emph{arXiv
  preprint arXiv:2110.15800} (2021) }.

\bibitem{bhat2022future}
P.~Bhat, S.~Jindariani, G.~Ambrosio, G.~Apollinari, S.~Belomestnykh, A.~Bross
  et~al., \emph{Future collider options for the {US}}, {\emph{arXiv preprint
  arXiv:2203.08088} (2022) }.

\bibitem{abramowicz2019physics}
H.~Abramowicz and R.~Forty, \emph{Physics briefing book [input for the
  {European Strategy for Particle Physics Update 2020}]},
  \href{https://arxiv.org/abs/1910.11775}{{\ttfamily 1910.11775}}.

\bibitem{blondel2022footprint}
A.~Blondel and P.~Janot, \emph{The environmental footprint of future {Higgs}
  boson studies: Who is the greenest?}, {\emph{arXiv preprint arXiv:2208.10466}
  (2022) }.

\bibitem{benedikt2019physics}
M.~Benedikt and F.~Zimmermann, \emph{The physics and technology of the {Future
  Circular Collider}}, {\emph{Nature Reviews Physics} {\bfseries 1} (2019)
  238}.

\bibitem{fcc}
{{The FCC Collaboration}}, \emph{{FCC Physics Opportunities: Future Circular
  Collider Conceptual Design Report Volume 1}}, {\emph{{European Physical
  Journal C}} {\bfseries 79} (2019) }.

\bibitem{fccee}
{{The FCC Collaboration}}, \emph{{FCC-ee: The Lepton Collider}},
  {\emph{{Eur.~Phys.~J.~Spec.~Top.}} {\bfseries 228} (2019) }.

\bibitem{council-fcc-1}
{CERN Council}, \emph{{Organisational structure of the FCC feasibility study.
  Restricted CERN Council - Two-Hundred-and-Third Session}},
  {\emph{CERN/SPC/1155/Rev.2} (2021) }.

\bibitem{council-fcc-2}
{CERN Council}, \emph{{{Main deliverables and timeline of the FCC feasibility
  study. Restricted CERN Council - Two-Hundred-and-Third Session}}},
  {\emph{CERN/SPC/1161} (2021) }.

\bibitem{FCCsnowmass}
I.~Agapov et~al., \emph{Future circular lepton collider {FCC-ee}: Overview and
  status}, {\emph{arXiv preprint arXiv:2203.08310} (2022) }.

\bibitem{HELEN}
S.~Belomestnykh, P.~Bhat, A.~Grassellino, M.~Checchin, D.~Denisov, R.~Geng
  et~al., \emph{{Higgs-Energy LEptoN (HELEN) Collider based on advanced
  superconducting radio frequency technology}}, {\emph{arXiv preprint
  arXiv:2203.08211} (2022) }.

\bibitem{litvinenko2020high}
V.N.~Litvinenko, T.~Roser and M.~Chamizo-Llatas, \emph{High-energy
  high-luminosity $e^+e^-$ collider using energy-recovery linacs},
  {\emph{Physics Letters B} {\bfseries 804} (2020) 135394}.

\bibitem{litvinenko2022cerc}
V.N.~Litvinenko, N.~Bachhawat, M.~Chamizo-Llatas, F.~Meot and T.~Roser,
  \emph{{CERC - circular $e^+e^-$ collider using Energy-Recovery Linac}},
  {\emph{arXiv preprint arXiv:2203.07358} (2022) }.

\bibitem{litvinenko2022relic}
V.N.~Litvinenko, N.~Bachhawat, M.~Chamizo-Llatas, Y.~Jing, F.~Meot,
  I.~Petrushina et~al., \emph{The {ReLiC}: Recycling linear $ e^+ e^-$
  collider}, {\emph{arXiv preprint arXiv:2203.06476} (2022) }.

\bibitem{telnov2021high}
V.~Telnov, \emph{A high-luminosity superconducting twin $e^+e^-$ linear
  collider with energy recovery}, {\emph{Journal of Instrumentation} {\bfseries
  16} (2021) P12025}.

\bibitem{CERN2022RnD}
D.~Newbold et~al., \emph{European strategy for particle physics accelerator
  r\&d roadmap},  Tech. Rep. CERN-2022-001 (2022).

\bibitem{barklow2022xcc}
T.~Barklow, S.~Dong, C.~Emma, J.~Duris, Z.~Huang, A.~Naji et~al., \emph{{XCC}:
  An {X-ray FEL}-based $\gamma \gamma$ collider {Higgs} factory}, {\emph{arXiv
  preprint arXiv:2203.08484} (2022) }.

\bibitem{AF3summary}
A.~Faus-Golfe, G.~Hoffstaetter, Q.~Qin and F.~Zimmermann, \emph{{Accelerators
  for Electroweak Physics and Higgs Boson Studies (Report of the Snowmass'21
  AF3)}}, {\emph{arXiv preprint arXiv:2209.05827} (2022) }.

\bibitem{bruning2019high}
O.~Bruning and L.~Rossi, \emph{The high-luminosity {Large Hadron Collider}},
  {\emph{Nature Reviews Physics} {\bfseries 1} (2019) 241}.

\bibitem{long2021muon}
K.R.~Long, D.~Lucchesi, M.A.~Palmer, N.~Pastrone, D.~Schulte and V.~Shiltsev,
  \emph{Muon colliders to expand frontiers of particle physics}, {\emph{Nature
  Physics} {\bfseries 17} (2021) 289}.

\bibitem{benedikt2022future}
M.~Benedikt, A.~Chance, B.~Dalena, D.~Denisov, M.~Giovannozzi, J.~Gutleber
  et~al., \emph{Future circular hadron collider {FCC-hh}: Overview and status},
  {\emph{arXiv preprint arXiv:2203.07804} (2022) }.

\bibitem{tang2022snowmass}
J.~Tang, Y.~Zhang, Q.~Xu, J.~Gao, X.~Lou and Y.~Wang, \emph{Snowmass 2021 white
  paper {AF4-SPPC}}, {\emph{arXiv preprint arXiv:2203.07987} (2022) }.

\bibitem{tang2022design}
J.~Tang, \emph{Design concept for a future {Super Proton-Proton Collider}},
  {\emph{Frontiers in Physics} (2022) 138}.

\bibitem{ESPPu}
{European Strategy Group}, \emph{{2020 Update of the European Strategy for
  Particle Physics (Brochure)}}, {\emph{CERN-ESU-015} (2020) }.

\bibitem{ferrario2021advanced}
M.~Ferrario and R.~Assmann, \emph{Advanced accelerator concepts}, {\emph{arXiv
  preprint arXiv:2103.10843} (2021) }.

\bibitem{jing2022continuous}
C.~Jing, J.~Power, J.~Shao, G.~Ha, P.~Piot, X.~Lu et~al., \emph{Continuous and
  coordinated efforts of structure wakefield acceleration {(SWFA)} development
  for an energy frontier machine}, {\emph{arXiv preprint arXiv:2203.08275}
  (2022) }.

\bibitem{AF4summary}
M.~Palmer et~al., \emph{{Report of the Snowmass'21 AF4}}, {\emph{arXiv preprint
  arXiv:2209.xxxxx} (2022) }.

\bibitem{eeFORUMsummary}
M.~Chamizo-Llatas, S.~Dasu, U.~Heitz et~al., \emph{{Report of the Snowmass'21
  $e+e-$ Collider Forum}}, {\emph{arXiv preprint arXiv:2209.03472} (2022) }.

\bibitem{mumuFORUMsummary}
K.~Black et~al., \emph{{Muon Collider Forum Report}}, {\emph{arXiv preprint
  arXiv:2209.01318} (2022) }.

\bibitem{IFC}
P.~Bhat et~al., \emph{{U.S. National Accelerator R\&D Program on Future
  Colliders}}, {\emph{arXiv preprint arXiv:2207.06213} (2022) }.

\bibitem{shiltsev2014cost}
V.~Shiltsev, \emph{A phenomenological cost model for high energy particle
  accelerators}, {\emph{Journal of Instrumentation} {\bfseries 9} (2014)
  T07002}.

\bibitem{gourlay2016us}
S.A.~Gourlay, S.O.~Prestemon, A.V.~Zlobin, L.~Cooley and D.~Larbalestier,
  \emph{{The US} magnet development program plan},  Tech. Rep. Lawrence
  Berkeley National Lab.(LBNL), Berkeley, CA (United States) (2016).

\bibitem{ambrosio2022strategic}
G.~Ambrosio, K.~Amm, M.~Anerella, G.~Apollinari, D.~Arbelaez, B.~Auchmann
  et~al., \emph{{A Strategic Approach to Advance Magnet Technology for Next
  Generation Colliders}}, {\emph{arXiv preprint arXiv:2203.13985} (2022) }.

\bibitem{vedrine2022high}
P.~V{\'e}drine, L.~Garcia-Tabar{\`e}s, B.~Auchmann, A.~Ballarino, B.~Baudouy,
  L.~Bottura et~al., \emph{High field magnet development for {HEP in Europe: a
  proposal from LDG HFM} expert panel}, {\emph{arXiv preprint arXiv:2203.08054}
  (2022) }.

\bibitem{alexahin2022critical}
Y.I.~Alexahin, E.~Barzi, E.~Gianfelice-Wendt, V.~Kapin, V.~Kashikhin, N.~Mokhov
  et~al., \emph{Critical problems of energy frontier muon colliders: optics,
  magnets and radiation}, {\emph{arXiv preprint arXiv:2203.10431} (2022) }.

\bibitem{piekarz2022fast}
H.~Piekarz, B.~Claypool, S.~Hays, M.~Kufer, V.~Shiltsev, A.~Zlobin et~al.,
  \emph{Fast cycling {HTS} based superconducting accelerator magnets:
  Feasibility study and readiness demonstration program driven by neutrino
  physics and muon collider needs}, {\emph{arXiv preprint arXiv:2203.06253}
  (2022) }.

\bibitem{ambrosio2022white}
G.~Ambrosio, G.~Apollinari, M.~Baldini, R.~Carcagno, C.~Boffo, B.~Claypool
  et~al., \emph{{White paper on leading-edge technology and
  feasibility-directed (LEAF) Program aimed at readiness demonstration for
  energy frontier circular colliders by the next decade}}, {\emph{arXiv
  preprint arXiv:2203.07654} (2022) }.

\bibitem{AF7MAGsummary}
S.~Izquierdo-Bermudez, G.~Sabbi and A.~Zlobin, \emph{{Accelerator Technology
  Magnets (AF07 Report)}}, {\emph{arXiv preprint arXiv:2208.13349} (2022) }.

\bibitem{grassellino2017unprecedented}
A.~Grassellino, A.~Romanenko, Y.~Trenikhina, M.~Checchin, M.~Martinello,
  O.~Melnychuk et~al., \emph{Unprecedented quality factors at accelerating
  gradients up to {45 MV/m} in niobium superconducting resonators via low
  temperature nitrogen infusion}, {\emph{Superconductor Science and Technology}
  {\bfseries 30} (2017) 094004}.

\bibitem{nanni2022c}
E.A.~Nanni, M.~Breidenbach, C.~Vernieri, S.~Belomestnykh, P.~Bhat, S.~Nagaitsev
  et~al., \emph{C$^3$ demonstration research and development plan},
  {\emph{arXiv preprint arXiv:2203.09076} (2022) }.

\bibitem{ives2022high}
R.L.~Ives, M.~Read, T.~Bui, D.~Marsden, G.~Collins, B.~Chase et~al., \emph{High
  efficiency, low cost, {RF} sources for accelerators and colliders},
  {\emph{arXiv preprint arXiv:2203.12043} (2022) }.

\bibitem{AF7RFsummary}
S.~Belomestnykh et~al., \emph{{RF Accelerator Technology R\&D: Report of AF7-RF
  Topical Group to Snowmass 2021}}, {\emph{arXiv preprint arXiv:2208.12368}
  (2022) }.

\bibitem{ammigan2022novel}
K.~Ammigan, S.~Bidhar, F.~Pellemoine, V.~Pronskikh, D.~Pushka, K.~Yonehara
  et~al., \emph{Novel materials and concepts for next-generation high power
  target applications}, {\emph{arXiv preprint arXiv:2203.08357} (2022) }.

\bibitem{barbier2022modeling}
C.~Barbier, S.~Bidhar, M.~Calviani, J.~Dooling, J.~Gao, A.~Jacques et~al.,
  \emph{Modeling needs for high power targets}, {\emph{arXiv preprint
  arXiv:2203.04714} (2022) }.

\bibitem{yonehara2022radiation}
K.~Yonehara, \emph{Radiation hardened beam instrumentations for multi-mega-watt
  beam facilities}, {\emph{arXiv preprint arXiv:2203.06024} (2022) }.

\bibitem{pellemoine2022irradiation}
F.~Pellemoine, C.~Barbier, Y.~Sun, K.~Ammigan, S.~Bidhar, B.~Zwaska et~al.,
  \emph{Irradiation facilities and irradiation methods for high power targets},
  {\emph{arXiv preprint arXiv:2203.08239} (2022) }.

\bibitem{chaikovska2022positron}
I.~Chaikovska, R.~Chehab, V.~Kubytskyi, S.~Ogur, A.~Ushakov, A.~Variola et~al.,
  \emph{Positron sources: from conventional to advanced accelerator
  concepts-based colliders}, {\emph{arXiv preprint arXiv:2202.04939} (2022) }.

\bibitem{AF7TSsummary}
F.~Pellemoine, C.~Barbier and Y.~Sun, \emph{{ Snowmass2021 Topical Group Report
  from AF07: Targets and Sources }}, {\emph{arXiv preprint arXiv:2208.13641}
  (2022) }.

\bibitem{litos20169}
M.~Litos, E.~Adli, J.~Allen, W.~An, C.~Clarke, S.~Corde et~al., \emph{{9 GeV
  energy gain in a beam-driven plasma wakefield accelerator}}, {\emph{Plasma
  Physics and Controlled Fusion} {\bfseries 58} (2016) 034017}.

\bibitem{gonsalves2019petawatt}
A.~Gonsalves, K.~Nakamura, J.~Daniels, C.~Benedetti, C.~Pieronek, T.~De~Raadt
  et~al., \emph{Petawatt laser guiding and electron beam acceleration to {8
  GeV} in a laser-heated capillary discharge waveguide}, {\emph{Physical review
  letters} {\bfseries 122} (2019) 084801}.

\bibitem{adli2018acceleration}
E.~Adli, A.~Ahuja, O.~Apsimon, R.~Apsimon, A.-M.~Bachmann, D.~Barrientos
  et~al., \emph{Acceleration of electrons in the plasma wakefield of a proton
  bunch}, {\emph{Nature} {\bfseries 561} (2018) 363}.

\bibitem{assmann2020eupraxia}
R.~Assmann, M.~Weikum, T.~Akhter, D.~Alesini, A.~Alexandrova, M.~Anania et~al.,
  \emph{Eupraxia conceptual design report}, {\emph{The European Physical
  Journal Special Topics} {\bfseries 229} (2020) 3675}.

\bibitem{wang2021fel}
W.~Wang, K.~Feng, L.~Ke, C.~Yu, Y.~Xu, R.~Qi et~al., \emph{Free-electron lasing
  at 27 nanometres based on a laser wakefield accelerator}, {\emph{Nature}
  {\bfseries 595} (2021) 516}.

\bibitem{shpakov2021first}
V.~Shpakov, M.~Anania, M.~Behtouei, M.~Bellaveglia, A.~Biagioni, M.~Cesarini
  et~al., \emph{First emittance measurement of the beam-driven plasma wakefield
  accelerated electron beam}, {\emph{Physical Review Accelerators and Beams}
  {\bfseries 24} (2021) 051301}.

\bibitem{gai1988experimental}
W.~Gai, P.~Schoessow, B.~Cole, R.~Konecny, J.~Norem, J.~Rosenzweig et~al.,
  \emph{Experimental demonstration of wake-field effects in dielectric
  structures}, {\emph{Physical review letters} {\bfseries 61} (1988) 2756}.

\bibitem{gao2008design}
F.~Gao, M.~Conde, W.~Gai, C.~Jing, R.~Konecny, W.~Liu et~al., \emph{Design and
  testing of a {7.8 GHz} power extractor using a cylindrical dielectric-loaded
  waveguide}, {\emph{Physical Review Special Topics-Accelerators and Beams}
  {\bfseries 11} (2008) 041301}.

\bibitem{england2022laser}
R.~England, D.~Filippetto, G.~Torrisi, A.~Bacci, G.~Della~Valle, D.~Mascali
  et~al., \emph{Laser-driven structure-based accelerators}, {\emph{arXiv
  preprint arXiv:2203.08981} (2022) }.

\bibitem{AF6summary}
C.~Geddes, M.~Hogan, P.~Musumeci and R.~Assmann, \emph{{Report of Snowmass 21
  Accelerator Frontier Topical Group 6 on Advanced Accelerators}}, {\emph{arXiv
  preprint arXiv:2208.13279} (2022) }.

\bibitem{fedotov2020experimental}
A.~Fedotov, Z.~Altinbas, S.~Belomestnykh, I.~Ben-Zvi, M.~Blaskiewicz,
  M.~Brennan et~al., \emph{Experimental demonstration of hadron beam cooling
  using radio-frequency accelerated electron bunches}, {\emph{Physical Review
  Letters} {\bfseries 124} (2020) 084801}.

\bibitem{jarvis2022first}
J.~Jarvis, V.~Lebedev, A.~Romanov, D.~Broemmelsiek, K.~Carlson,
  S.~Chattopadhyay et~al., \emph{First experimental demonstration of optical
  stochastic cooling}, {\emph{Nature} {\bfseries 608} (2022) 2870292}.

\bibitem{duris2020bayesian}
J.~Duris, D.~Kennedy, A.~Hanuka, J.~Shtalenkova, A.~Edelen, P.~Baxevanis
  et~al., \emph{Bayesian optimization of a free-electron laser},
  {\emph{Physical review letters} {\bfseries 124} (2020) 124801}.

\bibitem{arpaia2021machine}
P.~Arpaia, G.~Azzopardi, F.~Blanc, G.~Bregliozzi, X.~Buffat, L.~Coyle et~al.,
  \emph{Machine learning for beam dynamics studies at the {CERN Large Hadron
  Collider}}, {\emph{Nuclear Instruments and Methods in Physics Research
  Section A: Accelerators, Spectrometers, Detectors and Associated Equipment}
  {\bfseries 985} (2021) 164652}.

\bibitem{edelen2016neural}
A.L.~Edelen, S.G.~Biedron, B.E.~Chase, D.~Edstrom, S.V.~Milton and P.~Stabile,
  \emph{Neural networks for modeling and control of particle accelerators},
  \href{https://doi.org/10.1109/TNS.2016.2543203}{\emph{IEEE Transactions on
  Nuclear Science} {\bfseries 63} (2016) 878}.

\bibitem{edelen2016first}
A.~Edelen, S.~Biedron, S.~Milton and J.~Edelen, \emph{First steps toward
  incorporating image based diagnostics into particle accelerator control
  systems using convolutional neural networks}, {\emph{arXiv preprint
  arXiv:1612.05662} (2016) }.

\bibitem{koser2021input}
D.~Koser, L.~Waites, D.~Winklehner, M.~Frey, A.~Adelmann and J.~Conrad,
  \emph{Input beam matching and beam dynamics design optimization of the
  {IsoDAR RFQ} using statistical and machine learning techniques}, {\emph{arXiv
  preprint arXiv:2112.02579} (2021) }.

\bibitem{ariniello2019transverse}
R.~Ariniello, C.~Doss, K.~Hunt-Stone, J.~Cary and M.~Litos, \emph{Transverse
  beam dynamics in a plasma density ramp}, {\emph{Physical Review Accelerators
  and Beams} {\bfseries 22} (2019) 041304}.

\bibitem{lindstrom2021staging}
C.A.~Lindstr{\o}m, \emph{Staging of plasma-wakefield accelerators},
  {\emph{Physical Review Accelerators and Beams} {\bfseries 24} (2021) 014801}.

\bibitem{scandale2019channeling}
W.~Scandale and A.~Taratin, \emph{{Channeling and volume reflection of
  high-energy charged particles in short bent crystals. Crystal assisted
  collimation of the accelerator beam halo}}, {\emph{Physics Reports}
  {\bfseries 815} (2019) 1}.

\bibitem{shiltsev2021electronlens}
V.~Shiltsev, \emph{Electron lenses: historical overview and outlook},
  {\emph{Journal of Instrumentation} {\bfseries 16} (2021) P03039}.

\bibitem{danilov2010nonlinear}
V.~Danilov and S.~Nagaitsev, \emph{Nonlinear accelerator lattices with one and
  two analytic invariants}, {\emph{Physical Review Special Topics-Accelerators
  and Beams} {\bfseries 13} (2010) 084002}.

\bibitem{nagaitsev2021accelerator}
S.~Nagaitsev, Z.~Huang, J.~Power, J.-L.~Vay, P.~Piot, L.~Spentzouris et~al.,
  \emph{Accelerator and beam physics research goals and opportunities},
  {\emph{arXiv preprint arXiv:2101.04107} (2021) }.

\bibitem{antipov2017iota}
S.~Antipov, D.~Broemmelsiek, D.~Bruhwiler, D.~Edstrom, E.~Harms, V.~Lebedev
  et~al., \emph{{IOTA (Integrable Optics Test Accelerator): facility and
  experimental beam physics program}}, {\emph{Journal of Instrumentation}
  {\bfseries 12} (2017) T03002}.

\bibitem{AF1summary}
M.~Bai, Z.~Huang and S.~Lund, \emph{{Summary Report of AF1 to Snowmass 2021:
  Beam Physics and Accelerator Education within the Accelerator Frontier}},
  {\emph{arXiv preprint arXiv:2209.07668} (2022) }.

\bibitem{ABPGC}
S.~Nagaitsev et~al., \emph{{Accelerator and Beam Physics: Grand Challenges and
  Research Opportunities}}, {\emph{arXiv preprint arXiv:2203.06811} (2022) }.

\end{thebibliography}\endgroup

\end{document}